\documentclass[twocolumn]{aastex701}
\usepackage{xcolor}
\usepackage{amsmath}
\usepackage{lipsum}
\usepackage{comment}
\usepackage{makecell}
\usepackage{hyperref}
\usepackage{csvsimple}
\usepackage{xspace}
\usepackage{longtable,array,booktabs}
\usepackage{subcaption}

\newcommand{\msun}{\,\mathrm{M}_\odot}

\defcitealias{2026ApJ...996...69M}{M26}
\newcommand{\miller}{\citetalias{2026ApJ...996...69M}\xspace}

\begin{document}

\title{Atypical White Dwarfs in Open Clusters}

\correspondingauthor{David R. Miller}
\email{drmiller@phas.ubc.ca}

\author[0000-0002-4591-1903]{David R. Miller}
\affiliation{Department of Physics and Astronomy, University of British Columbia, Vancouver, BC V6T 1Z1, Canada}
\email{drmiller@phas.ubc.ca}

\author[0000-0002-4770-5388]{Ilaria Caiazzo}
\affiliation{Institute of Science and Technology Austria, Am Campus 1, Klosterneuburg, 3400, Austria}
\email{Ilaria.Caiazzo@ist.ac.at}

\author[0000-0001-9739-367X]{Jeremy Heyl}
\affiliation{Department of Physics and Astronomy, University of British Columbia, Vancouver, BC V6T 1Z1, Canada}
\email{heyl@phas.ubc.ca}

\author[0000-0001-9873-0121]{Pier-Emmanuel Tremblay}
\affiliation{Department of Physics, University of Warwick, Coventry, CV4 7AL, UK}
\email{P.Tremblay@warwick.ac.uk}


\begin{abstract}

Open clusters are powerful laboratories for investigating the formation and evolution of white dwarfs (WDs), as their well-constrained ages link WDs to their progenitor stars. This is particularly valuable for atypical WDs, including magnetic WDs and those with non-hydrogen-dominated spectra, whose origins can be difficult to disentangle in the field. We assemble the largest sample to date of atypical WDs in open clusters: eight non-magnetic WDs with helium-dominated spectra, five with featureless spectra, and five magnetic WDs showing hydrogen or helium lines. We find no significant difference between the initial--final mass relations of hydrogen- and helium-atmosphere WDs, nor robust spatial or kinematic differences. The strongest difference from the field is a deficit of non-DAs between $0.8$ and $1.0\msun$, where the field predicts $6.0^{+0.8}_{-0.7}$ WDs, yet none are observed, with a probability of only $0.37\%$. Massive magnetic WDs show a similar deficit: only one of 20 cluster WDs with $M_{\mathrm{WD}}\geq1.0\msun$ and $T_{\mathrm{eff}}\geq20{,}000$~K is magnetic, compared with 10 of 16 in the field. Together, these deficits are consistent with an additional field population of massive merger remnants, particularly high-mass magnetic WDs and warm/hot DQs. Such objects may be underrepresented in clusters because young systems provide limited time for certain classes of delayed mergers, while older clusters lose members or dissolve; the field accumulates binary evolution products over Galactic timescales. Magnetic cluster WDs consistent with single-star evolution further suggest mass-dependent magnetic-field emergence, with fields appearing early in some massive WDs while young lower-mass WDs remain consistent with delayed emergence.

\end{abstract}

\keywords{ \uat{Open star clusters}{1160} --- \uat{Stellar evolution}{1599} --- \uat{Stellar magnetic fields}{1610} --- \uat{Stellar mergers}{2157} --- \uat{White dwarf stars}{1799}}


\section{Introduction}
\label{sec:intro}

A large fraction of white dwarfs (WDs) have spectra dominated by hydrogen lines and are classified as DAs. Other spectral classes, collectively termed non-DAs, show helium, carbon, or metal features, or have featureless spectra. Magnetic fields are found across several spectral classes, and the relative incidence of different spectral types and of magnetism varies strongly with WD mass and effective temperature, reflecting the effects of both progenitor history and subsequent cooling evolution \citep[see][]{2005MNRAS.356.1576W,2021MNRAS.507.5902B,2024MNRAS.527.8687O,2024Ap&SS.369...43B,2025ApJ...990...25M}. This is particularly apparent at high WD masses, where field samples contain substantial populations of magnetic and otherwise peculiar WDs \citep{2003AJ....125..348L,2019ApJ...885...74C,2024ApJ...974...12J}.

Several physical processes may contribute to this diversity. Diffusion and convective mixing can alter the atmospheric composition as a WD cools \citep{2008ApJ...672.1144T,2018ApJ...857...56R,2020MNRAS.492.3540C}. Magnetism, meanwhile, may arise through several channels, including inheritance from the progenitor, crystallization-driven dynamos, and binary interactions or mergers \citep{2004MNRAS.355L..13T,2004IAUS..224..879K,2005MNRAS.356.1576W,2008MNRAS.387..897T,2011PNAS..108.3135N,2012ApJ...749...25G,2014MNRAS.437..675W,2015SSRv..191..111F,2015Sci...350..423F,2016Natur.529..364S,2016ApJ...824...14C,2017ApJ...836L..28I,2022MNRAS.514.4111G,2024A&A...691L..21C,2026A&A...708L..14E}. Recent field studies suggest that the relative importance of these channels varies across the WD population, with older, lower-mass magnetic WDs broadly consistent with single-star origins, while merger formation may contribute more prominently among young, massive magnetic WDs \citep{2021MNRAS.507.5902B,2022ApJ...935L..12B,2025ApJ...990...25M}. Separating these channels in the field remains difficult because the total ages and progenitor properties of individual WDs are generally unknown, and the observed WD properties such as WD mass, cooling age relative to kinematics, magnetic field strength, rotation period, and atmospheric composition may not uniquely separate a single-star remnant from a merger product \citep{2020MNRAS.499.1890M,2020ApJ...898...84K,2026ApJ...999..146O}.

Star clusters provide an important advantage for separating these evolutionary pathways. Because cluster members are approximately coeval, the cluster age can be combined with the WD cooling age to infer the progenitor lifetime and mass, providing the basis for cluster WD initial--final mass relation (IFMR) studies \citep[e.g.][]{2005MNRAS.361.1131F,2018ApJ...866...21C,2021ApJ...912..165R,2026ApJ...996...69M}. When astrometry supports cluster membership but the WD properties are inconsistent with single-star evolution at the cluster age, the object can be identified as a candidate product of binary evolution \citep{2004MNRAS.355L..39D,2018ApJ...867...62W,2025ApJ...979L...1L,2026ApJ...999..167Y,2026arXiv260711299J}. The cluster age can then constrain the possible binary history by limiting the available evolutionary time \citep{2014ApJ...783L...8G,2019ApJ...885...45G,2024ApJ...976..102G,2026arXiv260720611G}. For magnetic WDs, the cooling age places an upper limit on the time between WD formation and the appearance of a detectable surface magnetic field, while the cluster age and WD mass can test whether a merger origin is viable \citep{2020ApJ...901L..14C}.

These advantages have so far been limited by the small number of known non-DA and magnetic cluster WDs. Early studies found fewer helium-atmosphere WDs in young clusters than expected from field samples \citep{2005ApJ...618L.129K,2009ApJ...705..398D}. More recently, \citet{2021AJ....162..162B} compiled the known non-DA cluster population and found a possible tendency toward lower final masses than DAs at fixed progenitor mass. Magnetic cluster WDs have also provided clues to their evolutionary origins, with \citet{2020ApJ...901L..14C} suggesting that intermediate-mass progenitors may be more likely to produce magnetic WDs. Together, these studies point to potentially important differences between cluster and field populations, but the small number of known objects has limited comparisons.

The known cluster WD population has grown considerably in recent years. Gaia astrometry \citep{2016A&A...595A...1G,2018A&A...616A...1G,2021A&A...649A...1G,2023A&A...674A...1G} has enabled all-sky catalogues that greatly improve the census of Milky Way open clusters \citep{2020A&A...640A...1C,2021MNRAS.504..356D,2023A&A...673A.114H}. Combined with systematic searches for cluster WDs, large public spectroscopic surveys, and dedicated follow-up programs, these catalogues have helped substantially increase the overall cluster WD sample \citep{2026ApJ...996...69M}, including the number of known non-DA and magnetic cluster WDs.

Here, we report the discovery of five new candidate non-DA or magnetic WDs associated with open clusters, identified in the archival spectroscopic census of \cite{Miller2026Surveys}, and analyze the known sample from the literature. We use the additional evolutionary constraints from cluster membership to examine what these populations reveal about WD evolution in clusters and the field. We assemble an extensive literature census, supplemented by candidate atypical WDs identified in the aforementioned census \citep{Miller2026Surveys}, and use this sample to compare the incidence and properties of non-DA and magnetic WDs in clusters with those in the field. For the non-DA population, we also test for differences in the form of the IFMR and in spatial and kinematic distributions. For the magnetic population, we examine proposed formation channels, including fields produced in single-star progenitors, crystallization-driven magnetism, and mergers. Together, these analyses constrain the evolution of atypical WDs and the processes that shape these populations in clusters and the field.

We describe the sample selection in Section~\ref{sec:sample}, analyze the cluster and WD samples in Section~\ref{sec:analysis}, examine the non-DA WD population in Section~\ref{sec:nonDAs} and the magnetic WD population in Section~\ref{sec:magnetic_wds}, and summarize our findings in Section~\ref{sec:conclusions}.


\section{Sample Selection}
\label{sec:sample}

In this work, we define atypical cluster WDs as cluster-associated WDs that fall outside the dominant population of non-magnetic DAs. This includes magnetic WDs of any atmospheric type and non-magnetic non-DA WDs. Relevant spectral classes include WDs with spectra dominated by neutral helium lines (DBs), those with additional weaker hydrogen lines (DBAs), WDs with featureless spectra (DCs), and those showing carbon lines (DQs). We focus on isolated WDs and WDs in non-interacting systems for which the WD can be characterized independently. We only consider sources whose classification is confirmed, or at least strongly suggested, by spectroscopy. We exclude WDs in globular clusters, but consider moving groups where they define a sufficiently coherent color-magnitude diagram (CMD) and their ages are reasonably well constrained. 

We use the non-DA cluster WD compilation of \citet{2021AJ....162..162B} as the starting point for the literature component of the sample. Their search identified 13 non-DA WDs in open-cluster fields, of which seven were assessed as likely cluster members. We follow their membership assessment, which uses Gaia DR3 astrometry where available and alternative proper-motion information otherwise. We retain their viable cluster members, including the Hyades DBA WD EGGR~316 \citep{1976ApJ...210..524G,1984AJ.....89..830E}, the magnetic DBH WD in M47 \citep{2019ApJ...880...75R}, and the M67 DB/DBA and DC WDs with proper-motion membership support \citep{2010A&A...513A..50B,2018ApJ...867...62W,2021AJ....161..169C}, while excluding WD candidates from M34, M35, and NGC~6633 that lack sufficient membership support \citep{2006ApJ...643L.127W,2007AJ....133.1490W,2008AJ....135.2163R,2009ApJ...693..355W}. We retain the M47 DBH WD despite the infrared excess possibly associated with a late-type companion, as the WD is spectroscopically well characterized and the system shows no evidence of interaction \citep{2019ApJ...880...75R}.

We also include additional non-magnetic, non-DA WDs with literature cluster associations: the DC WDs Praesepe WD12 and Hyades LB~228 \citep{2026ApJ...996...69M}, the DB Praesepe~WD13 \citep{2019A&A...627A...4R,2025A&A...694A.258R,2026ApJ...996...69M}, the Ruprecht~147 DB R147-WD02 \citep{2020NatAs...4.1102M}, and the DBA IC~4756~WD1 \citep{2026ApJ...996...69M,2026ApJ..1007..183M}. Several magnetic WDs with established cluster associations are also included from the literature. These include the well-studied Praesepe DAH EGGR~59 \citep{2001ApJ...563..987C,2009MNRAS.395.1795C,2019MNRAS.483.3098S,2019A&A...627A...4R,2019MNRAS.486.5405G}, the ASCC~47 and M39 DAH WDs \citep{2020ApJ...901L..14C,2021ApJ...912..165R}, and the Stock~2 DAH WD \citep{2026ApJ...996...69M}.

To supplement the literature sample, we use candidate atypical cluster WDs identified in the spectroscopic census of \citet{Miller2026Surveys}. In that work, high-probability WDs from \citet{2021MNRAS.508.3877G}, corresponding to $P_{\rm WD}>0.75$, were first identified among the full membership lists of the \citet{2024A&A...686A..42H} cluster catalog and the Unified Cluster Catalog (UCC; \citealt{2023MNRAS.526.4107P}). These candidates were then crossmatched with the most recent public releases of three large low-resolution spectroscopic surveys: the Dark Energy Spectroscopic Instrument (DESI) DR1 \citep{2016arXiv161100036D,2026AJ....171..285D}, the Large Sky Area Multi-Object Fiber Spectroscopic Telescope (LAMOST) DR11 \citep{2012RAA....12.1197C}, and the Sloan Digital Sky Survey (SDSS) DR20 \citep{2000AJ....120.1579Y,2026arXiv260726149S}.

Candidate matches were then inspected spectroscopically, and only objects whose atypical classification was confirmed or strongly suggested by the available spectra were retained for this work. This inspection identified five additional atypical WDs not already included in the literature sample: three featureless WDs associated with HSC~453, HSC~1224, and HSC~1555, a DB associated with HSC~1542, and a magnetic WD associated with Alessi~13. The HSC~453 and HSC~1542 candidates are excluded because their proposed hosts lack coherent CMD sequences, preventing any reasonable age estimate. The HSC~1555 candidate is also excluded because no candidate cluster members are found sufficiently close to the main-sequence turnoff (MSTO) to meaningfully constrain the cluster age. This leaves HSC~1224 and Alessi~13 as cluster member candidates for further analysis. The census also independently recovered several objects already included in the sample, including EGGR~59, EGGR~316, M67:WD21, and Praesepe WDs~12 and 13.

The initial sample is summarized in Table~\ref{tab:wds_astrometry_type}. It contains 19 candidate atypical cluster WDs consisting of eight non-magnetic DB/DBA WDs, five DC WDs, and six magnetic DA/DB WDs showing Zeeman splitting. To avoid ambiguity with other literature works, we assign each object a standardized designation in the form Cluster-WDJHHMMSS$\pm$DDMMSS, based on Gaia DR3 coordinates where available and otherwise on the default ICRS coordinates at epoch J2000.0 listed on SIMBAD. We use these designations throughout the remainder of the paper. The membership column summarizes the astrometric evidence linking each WD to its proposed host cluster, distinguishing sources with Gaia astrometry consistent with cluster membership from those whose association is based on non-Gaia astrometry. Final inclusion in the cluster sample is assessed through the WD analysis in Section~\ref{sec:analysis}.

\begin{table*}[ht]
\tiny
\centering
\caption{Candidate atypical WD cluster members with select Gaia DR3 data and spectral type.} 
\label{tab:wds_astrometry_type}
\begin{tabular}{llcccccccc}
\toprule
Name &
Literature name &
Spectral type &
Gaia DR3 source ID &
RA &
Dec &
Plx &
pmRA &
pmDec &
Membership \\
&
&
&
&
[deg] &
[deg] &
[mas] &
[mas/yr] &
[mas/yr] &
\\
\midrule
\multicolumn{10}{l}{\textit{non-magnetic DB/DBA}} \\
\midrule
Hyades-WDJ044023+135845  & EGGR 316 & DBA & 3308403897837092992 & 70.100 & 13.979 & 21.76 & 95.29 & -20.84  & Gaia \\
IC4756-WDJ183724+053019 & IC 4756 WD1 & DBA & 4284010735661963648 & 279.352 & 5.505 & 2.2 & 1.2 & -5.6  & Gaia \\
M67-WDJ085029+115249 & M67:WD1 & DB & 604963941987041280 & 132.623 & 11.880 & 3.8 & -10.8 & -2.3  & Gaia \\
M67-WDJ085122+115329 & M67:WD18 & DBA & \ldots & \ldots & \ldots & \ldots & \ldots & \ldots  & Other \\
M67-WDJ085134+114217 & M67:WD21 & DB & 604898490980773888 & 132.895 & 11.705 & 1.0 & -11.5 & -3.4  & Gaia \\
M67-WDJ085145+114834 & M67:WD30 & DB & 604917079598643712 & 132.941 & 11.810 & \ldots & \ldots & \ldots & Other \\
Praesepe-WDJ081656+204945    & Praesepe WD13 & DB & 675767959626673152 & 124.234 & 20.829 & 4.79 & -29.79 & -13.70  & Gaia \\
Ruprecht147-WDJ191252-161435 & R147-WD02 & DB & 4088108859141437056 & 288.219 & -16.243 & 3.5 & -1.0 & -26.7  & Gaia \\
\midrule
\multicolumn{10}{l}{\textit{DC}} \\
\midrule
HSC1224-WDJ092525+525651 & \ldots & DC & 1019891219101344256 & 141.356 & 52.948 & 7.83 & -57.3 & -39.06  & Gaia \\
Hyades-WDJ041309+185936 & LB~228 & DC & 48522895539002624 & 63.288 & 18.994 & 16.76 & 83.30 & -42.45  & Gaia \\
M67-WDJ085143+115143 & M67:WD27 & DC & \ldots & \ldots & \ldots & \ldots & \ldots & \ldots  & Other \\
M67-WDJ085148+114143 & M67:WD31 & DC & \ldots & \ldots & \ldots & \ldots & \ldots & \ldots  & Other \\
Praesepe-WDJ082547+174818 & Praesepe WD12 & DC & 662152977721471488 & 126.448 & 17.805 & 5.0 & -29.2 & -10.43  & Gaia \\
\midrule
\multicolumn{10}{l}{\textit{magnetic DA/DB}} \\
\midrule
Alessi13-WDJ033301-384025 & \ldots & DAH & 4853382867764646912 & 53.256 & -38.674 & 10.1 & 34.30 & -5.99 & Gaia \\
ASCC47-WDJ083224-393543 & ASCC 47 & DAH & 5529347562661865088 & 128.101 & -39.595 & 1.45 & -7.97 & 5.1  & Gaia \\
M39-WDJ212659+484455 & M39 & DAH & 2170776080281869056 & 321.749 & 48.749 & 3.4 & -7.0 & -19.6  & Gaia \\
M47-WDJ073606-145131 & M47 & DBH & 3029912407273360512 & 114.029 & -14.859 & 2.5 & -8.1 & 2.1  & Gaia \\
Praesepe-WDJ083945+200015 & EGGR~59 & DAH & 664325543977630464 & 129.940 & 20.004 & 5.54 & -36.66 & -13.46  & Gaia \\
Stock2-WDJ020725+592444 & Stock 2 WD2 & DAH & 507105143670906624 & 31.857 & 59.412 & 2.9 & 16.1 & -13.6  & Gaia \\
\bottomrule
\end{tabular}
\begin{minipage}{\linewidth}
\medskip
{\textbf{Notes}. Membership is supported by Gaia astrometry for entries marked Gaia and by other astrometric measurements for entries marked Other.}
\end{minipage}
\end{table*}


\section{Cluster and WD Characterization}
\label{sec:analysis}


\subsection{Cluster Parameters}
\label{subsec:clusters}

\begin{figure}[ht]
    \centering
    \hspace*{-0.07\columnwidth}
    \includegraphics[width=1.0\columnwidth]{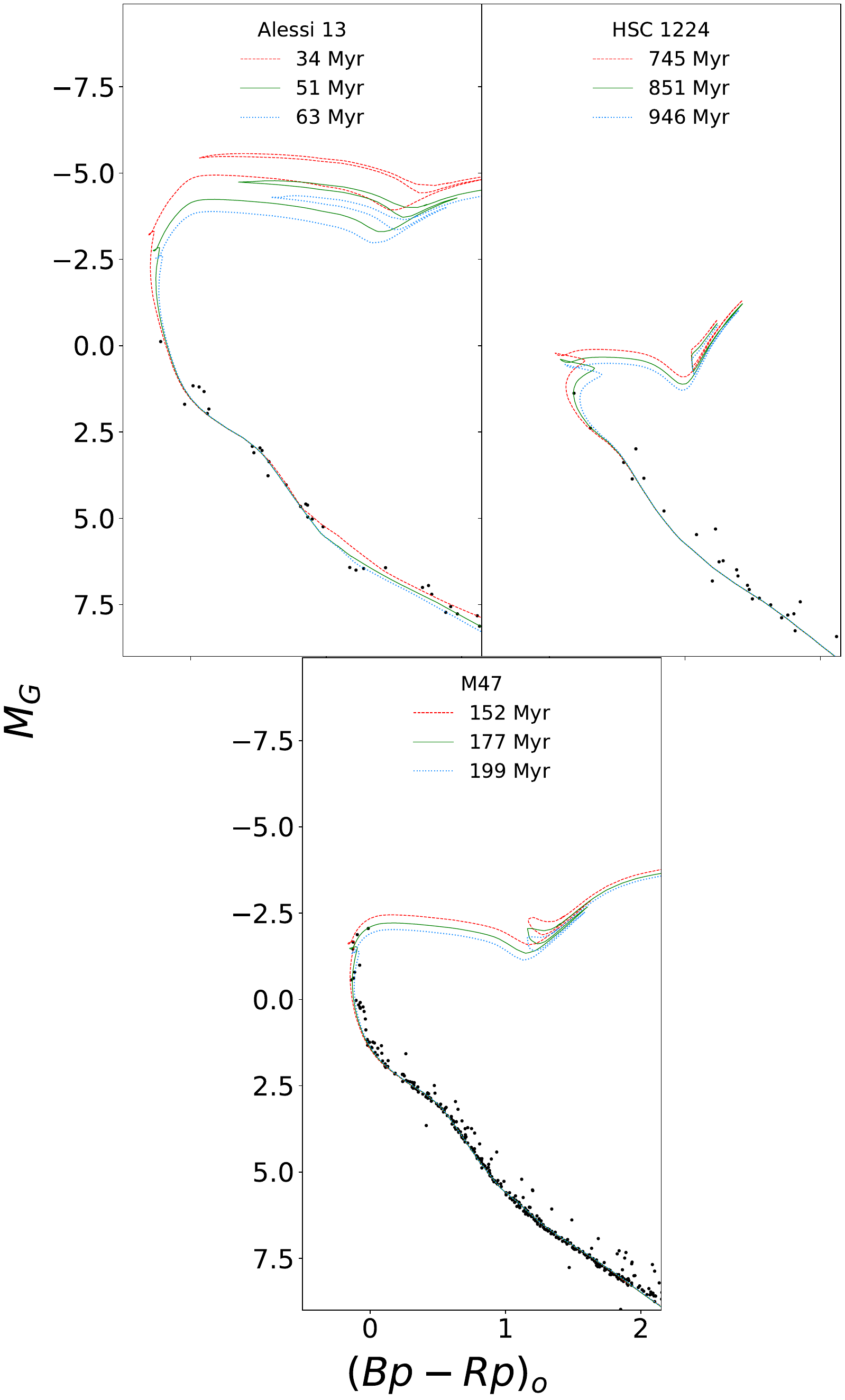}
    \caption{CMDs of Alessi~13 (top left), HSC~1224 (top right), and M47 (bottom) constructed from stars with \citet{2024A&A...686A..42H} membership probabilities $\geq50\%$. The adopted PARSEC isochrone from our heuristic fitting procedure is shown in green, with the lower and upper age bounds shown in red and blue, respectively.} 
    \label{fig:cluster_ages} 
\end{figure}

We determine cluster parameters using the heuristic MSTO-based PARSEC \citep{2012MNRAS.427..127B,2014MNRAS.445.4287T,2014MNRAS.444.2525C,2018MNRAS.476..496F,2022A&A...665A.126N} isochrone fitting procedure developed by \citet{2026ApJ...996...69M} (hereafter \miller). The best fit is selected using the MSTO, upper main sequence, and evolved sequence where available, with weighted $\chi^2$ variations used to estimate the age bounds. We adopt previously determined cluster parameters from \miller and \citet{2026ApJ..1007..183M}, both of which use the same fitting procedure. Only three clusters in the present sample lacked fits from these works; for these clusters, we fit the \citet{2024A&A...686A..42H} member candidates with membership probabilities $\geq50\%$. The resulting fits are shown in Fig.~\ref{fig:cluster_ages}, while the adopted cluster parameters for all clusters in the sample are summarized in Table~\ref{tab:clusters}. As in \miller, ASCC~47 is not included in the \citet{2024A&A...686A..42H} catalog, so we adopt the literature age from \citet{2020ApJ...901L..14C}. The quoted uncertainties reflect the internal fitting bounds from the adopted procedure, and do not include systematic uncertainties from stellar models, unresolved binaries, or membership selection.

\begin{table}[ht]
\tiny
\caption{Properties for candidate host clusters.}
\label{tab:clusters}
\centering
\setlength{\tabcolsep}{3pt}
\begin{tabular}{cccccc}
\toprule
Name &
Distance &
$A_\mathrm{V}$ &
$Z$ &
Age &
Source 
\\
 &
[pc] & 
[mag] & 
& 
[Myr] &
\\
\midrule
Alessi 13    & 104 & 0.096 & 0.020 & $51^{+12}_{-17}$        & this work \\
ASCC 47      & \ldots  & \ldots    & 0.015 & $90\pm20$       & \citet{2020ApJ...901L..14C} \\
HSC 1224     & 145 & 0.128 & 0.012 & $851^{+95}_{-106}$      & this work \\
Hyades       & 47  & 0.000 & 0.024 & $646^{+85}_{-84}$       & \miller \\
IC 4756      & 467 & 0.360 & 0.013 & $1{,}067^{+97}_{-85}$   & \citet{2026ApJ..1007..183M}            \\
M39          & 296 & 0.015 & 0.015 & $407^{+32}_{-20}$       & \miller \\
M47          & 468 & 0.189 & 0.018 & $177^{+22}_{-25}$       & this work                   \\
M67          & 837 & 0.132 & 0.012 & $3{,}819^{+292}_{-86}$  & \miller \\
Praesepe     & 183 & 0.059 & 0.019 & $741^{+72}_{-30}$       & \miller \\
Ruprecht 147 & 303 & 0.113 & 0.018 & $2{,}754^{+408}_{-124}$ & \miller \\
Stock 2      & 370 & \ldots & 0.015 & $437^{+22}_{-16}$      & \miller \\
\bottomrule
\end{tabular}
\begin{minipage}{\linewidth}
\medskip
{\textbf{Notes}. Distances are adopted from \citet{2024A&A...686A..42H} where available; the remaining parameters are derived using the heuristic isochrone fitting technique from the given source. As ASCC~47 is not included in \citet{2024A&A...686A..42H}, we follow \miller and adopt its parameters from \citet{2020ApJ...901L..14C}. No single reddening value is reported for Stock~2 because of substantial differential reddening.}
\end{minipage}
\end{table}


\subsection{WD Parameters}
\label{subsec:wd_params}

We derive WD atmospheric and cooling parameters for the candidate atypical cluster WDs. Spectroscopy is used to establish the classification in all cases. For DB/DBA WDs we determine atmospheric parameters from line-profile fitting, while DC and magnetic WDs are characterized photometrically. For magnetic WDs, this avoids the strong degeneracy between Zeeman broadening and surface gravity in standard line-profile fits. The adopted parameters are summarized in Table~\ref{tab:wds_derived}.

\begin{table*}
\scriptsize
\centering
\caption{Adopted atmospheric, cooling, and progenitor parameters for candidate atypical cluster WDs.} 
\label{tab:wds_derived}
\begin{tabular}{lccccccc}
\toprule
Name &
Type &
$\log g$ &
$T_\textrm{eff}$ &
Radius & 
$M_\textrm{WD}$ &
$t_\textrm{cool}$ & 
$M_\textrm{prog}$
\\
&
&
[$\mathrm{cm}\,\mathrm{s}^{-2}$] &
[K] &
[$\mathrm{km}$] &
[$\msun$] & 
[$\mathrm{Myr}$] & 
[$\msun$]
\\
\midrule
\multicolumn{2}{l}{\textit{non-magnetic DB/DBA}} &
\multicolumn{2}{c}{\hrulefill\ \textit{fit parameters}\ \hrulefill} &
\multicolumn{4}{c}{} \\
\midrule
Hyades-WDJ044023+135845     & DBA  & $8.28\pm0.10$ & $15{,}800\pm300$ & $7{,}300\pm500$ & $0.76^{+0.07}_{-0.06}$ & $290\pm50$ & $3.3\pm0.3$ \\
IC4756-WDJ183724+053019          & DBA  & $8.20\pm0.06$ & $23{,}700\pm600$ & $7{,}800\pm300$ & $0.73\pm0.04$ & $51^{+11}_{-9}$ & $2.22\pm0.07$\\
M67-WDJ085029+115249              & DB   & $8.36\pm0.04$ & $27{,}200\pm700$ & $6{,}900\pm200$ & $0.83\pm0.03$ & $44^{+8}_{-7}$ & \ldots \\
M67-WDJ085122+115329            & DBA  & $8.0\pm0.2$ & $12{,}400\pm300$ & $8{,}900^{+1{,}400}_{-1{,}300}$ & $0.58^{+0.14}_{-0.12}$ & $360^{+150}_{-100}$ & $1.38^{+0.02}_{-0.04}$ \\
M67-WDJ085134+114217             & DB   & $7.85\pm0.06$ & $35{,}700\pm700$ & $10{,}200^{+500}_{-400}$ & $0.55\pm0.03$          & $5.7\pm0.03$                 & $1.342^{+0.009}_{-0.031}$ \\
M67-WDJ085145+114834             & DB   & $8.20\pm0.04$ & $26{,}700\pm700$ & $7{,}800\pm200$ & $0.73\pm0.03$ & $25^{+5}_{-4}$ & $1.343^{+0.010}_{-0.030}$  \\
Praesepe-WDJ081656+204945        & DB   & $8.02\pm0.03$ & $26{,}100\pm400$ & $8{,}890\pm180$ & $0.619\pm0.017$ & $17.8^{+1.2}_{-1.1}$ & $2.57^{+0.04}_{-0.08}$ \\
Ruprecht147-WDJ191252-161435            & DB   & $8.15\pm0.05$ & $16{,}380\pm160$ & $8{,}000\pm300$ & $0.68\pm0.03$ & $202^{+19}_{-18}$ & $1.58^{+0.02}_{-0.07}$ \\
\midrule
\multicolumn{3}{l}{\textit{DC}} &
\multicolumn{2}{c}{\hrulefill\ \textit{fit parameters}\ \hrulefill} &
\multicolumn{3}{c}{} \\
\midrule
HSC1224-WDJ092525+525651              & DC           & $8.18\pm0.03$ & $11{,}400\pm180$ & $7{,}810\pm160$ & $0.693\pm0.019$ & $600\pm30$ & $3.6^{+0.9}_{-0.4}$ \\
Hyades-WDJ041309+185936       & DC           & $8.220\pm0.014$ & $10{,}800\pm70$ & $7{,}580\pm80$ & $0.719\pm0.009$ & $739\pm16$ & \ldots \\
M67-WDJ085143+115143              & DC           & $7.32\pm0.14$ & $7{,}200\pm300$ & $12{,}900\pm1{,}000$ & $0.26^{+0.05}_{-0.04}$ & $820^{+100}_{-50}$ & \ldots  \\
M67-WDJ085148+114143             & DC           & \ldots & $8{,}200\pm500$ & $16{,}100\pm1{,}600$ & \ldots & \ldots & \ldots\\
\midrule
\multicolumn{3}{l}{\textit{magnetic WDs}} &
\multicolumn{2}{c}{\hrulefill\ \textit{fit parameters}\ \hrulefill} &
\multicolumn{3}{c}{} \\
\midrule
Alessi13-WDJ033301-384025$^\star$             & DAH & $8.27\pm0.02$ & $11{,}090\pm120$ & $7{,}390\pm110$ & $0.774\pm0.013$ & $790^{+30}_{-60}$ & \ldots \\
ASCC47-WDJ083224-393543              & DAH  & $8.40\pm0.11$ & $110{,}000\pm4{,}000^\ddagger$ & $7{,}200\pm700$ & $0.97\pm0.05$ & $0.09\pm0.02$ & $5.5^{+0.5}_{-0.4}$ \\
M39-WDJ212659+484455                   & DAH  & $8.45\pm0.02$ & $16{,}900\pm200$ & $6{,}500\pm110$ & $0.897\pm0.014$ & $307^{+12}_{-10}$ & $5.3^{+0.5}_{-0.6}$  \\
M47-WDJ073606-145131                  & DBH  & $8.80^{+0.05}_{-0.04}$ & $31{,}000\pm2{,}000$ & $4{,}800\pm200$ & $1.10\pm0.02$ & $98^{+25}_{-20}$ & $5.8^{+1.8}_{-0.8}$  \\
Praesepe-WDJ082547+174818        & DC$^\dagger$ & $8.969^{+0.018}_{-0.017}$ & $20{,}000\pm500$ & $4{,}060\pm70$ & $1.156\pm0.007$ & $550^{+50}_{-40}$ & \ldots \\
Praesepe-WDJ083945+200015    & DAH  & $8.243\pm0.018$ & $15{,}400\pm200$ & $7{,}600\pm100$ & $0.762\pm0.011$ & $289\pm14$ & $3.3^{+0.3}_{-0.2}$  \\
Stock2-WDJ020725+592444         & DAH  & $8.43\pm0.06$ & $20{,}100\pm1{,}700$ & $6{,}600\pm300$ & $0.89\pm0.04$ & $180^{+50}_{-40}$ & $3.6^{+0.3}_{-0.2}$   \\
\bottomrule
\end{tabular}
\begin{minipage}{\linewidth}
\medskip
{\textbf{Notes}. Quantities marked as fit parameters are the values directly optimized in the fitting procedure for each group; the remaining quantities are inferred from those fitted parameters and the adopted cluster properties. DBA abundances: IC4756-WDJ183724+053019 has $\log(\mathrm{H}/\mathrm{He})=-4.26\pm0.09$, Hyades-WDJ044023+135845 has $\log(\mathrm{H}/\mathrm{He})=-4.61\pm0.07$, while M67-WDJ085122+115329 $\log(\mathrm{H}/\mathrm{He})$ is fixed to $-5$, following \citet{2021AJ....162..162B}. For M67-WDJ085148+114143, the fitted $T_{\rm eff}$ and radius suggest a mass lower than the bottom value ($0.2\msun$) of the evolutionary model grid; we therefore do not report $\log g$, mass, or cooling age.} Objects without progenitor masses do not yield a meaningful single-star progenitor solution. $^\dagger$Spectroscopically classified as DC; interpreted as highly magnetic, with the DA SED solution adopted. $^\ddagger$ At the high temperature of ASCC47-WDJ083224-393543, the SED provides only a weak temperature constraint. We therefore adopt the spectroscopic temperature determined from the magnetic atmosphere models of \citet{2020ApJ...901L..14C} and combine it with our SED-derived radius to determine the remaining parameters. $^\star$ Ruled out as an Alessi~13 member because its cooling age substantially exceeds the cluster age.
\end{minipage}
\end{table*}

\subsubsection{Spectroscopic DB/DBA Fitting}
\label{subsubsec:db_dba_fitting}

\begin{figure*}
    \centering
    \includegraphics[width=0.8\textwidth]{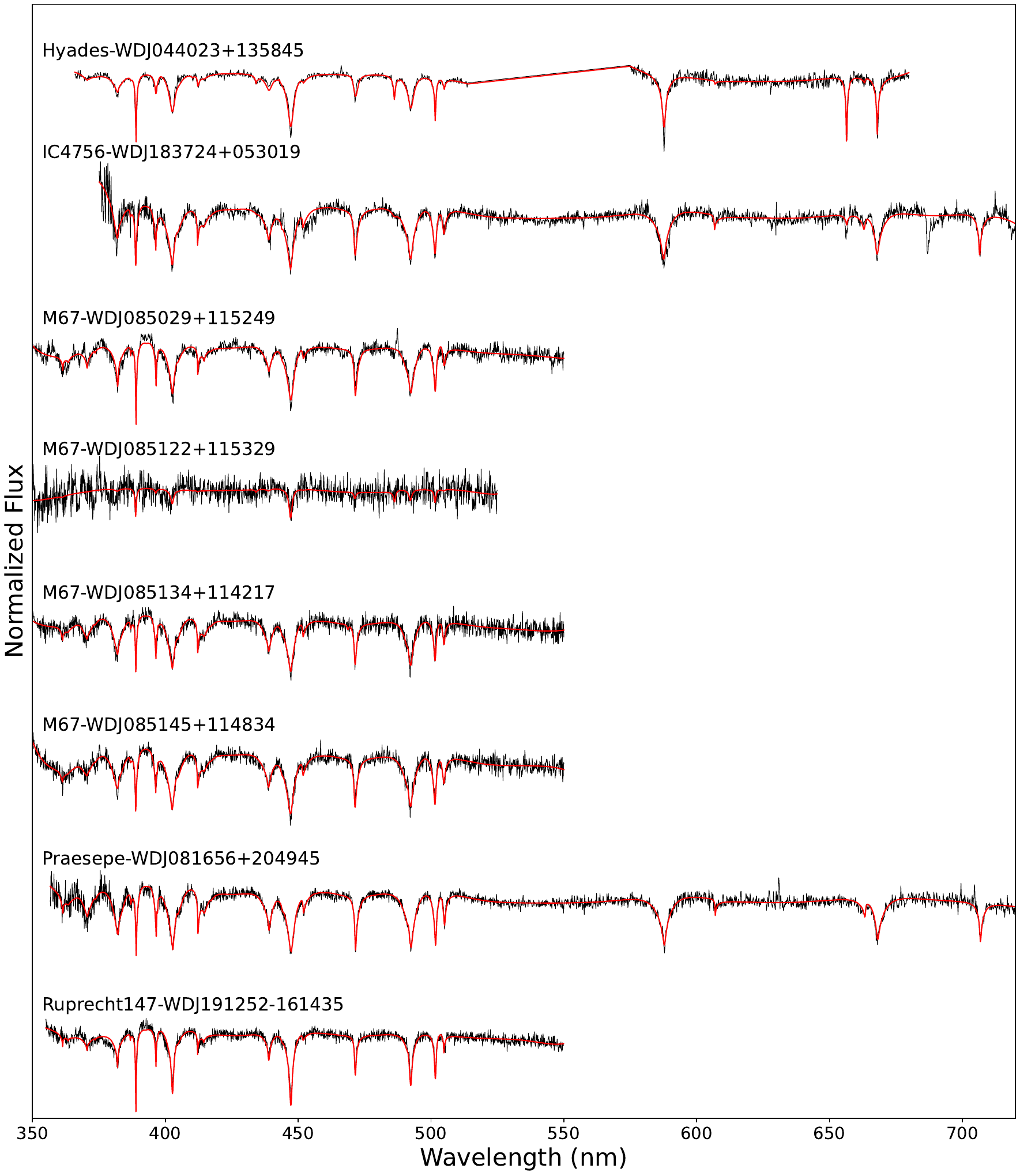}
    \caption{Normalized spectra of the candidate cluster DB and DBA WDs, with arbitrary vertical offsets for clarity. Observed spectra are shown in black with best-fit atmospheric models overlaid in red. Both are divided by a smooth continuum estimated from the best-fit model.}
    \label{fig:DB_spectra}
\end{figure*}

For the non-magnetic helium-atmosphere WDs, we fit the DB and DBA spectra with the model grid of \citet{2021MNRAS.501.5274C}, following the general line-profile fitting approach of \citet{2005ApJS..156...47L}. We first compare the observed spectrum to a grid of synthetic spectra together with a polynomial correction for the continuum shape. The spectrum is then normalized using fixed continuum points around the He~I lines, and the line profiles are fit simultaneously with a Levenberg--Marquardt least-squares routine. For DB WDs, this directly gives the best-fitting $T_\textrm{eff}$ and $\log g$. For DBA WDs, the hydrogen abundance must also be determined because even weak hydrogen can affect the atmospheric structure and the helium-line solution. We therefore use an iterative procedure following \citet{2011ApJ...737...28B}: for a fixed $\log(\mathrm{H}/\mathrm{He})$, we fit the He~I lines for $T_\textrm{eff}$ and $\log g$, then hold those values fixed and fit H$\alpha$ for $\log(\mathrm{H}/\mathrm{He})$. This process is repeated until the abundance changes by less than 0.05~dex. We then apply the 3D corrections of \citet{2021MNRAS.501.5274C} to the final fitted $T_\textrm{eff}$ and $\log g$ values, expect for M67-WDJ085134+114217, which is hotter than the upper temperature limit of the correction grid. The normalized spectra and best-fit models are shown in Fig.~\ref{fig:DB_spectra}.

\begin{figure*}
    \centering
    \includegraphics[width=0.8\textwidth]{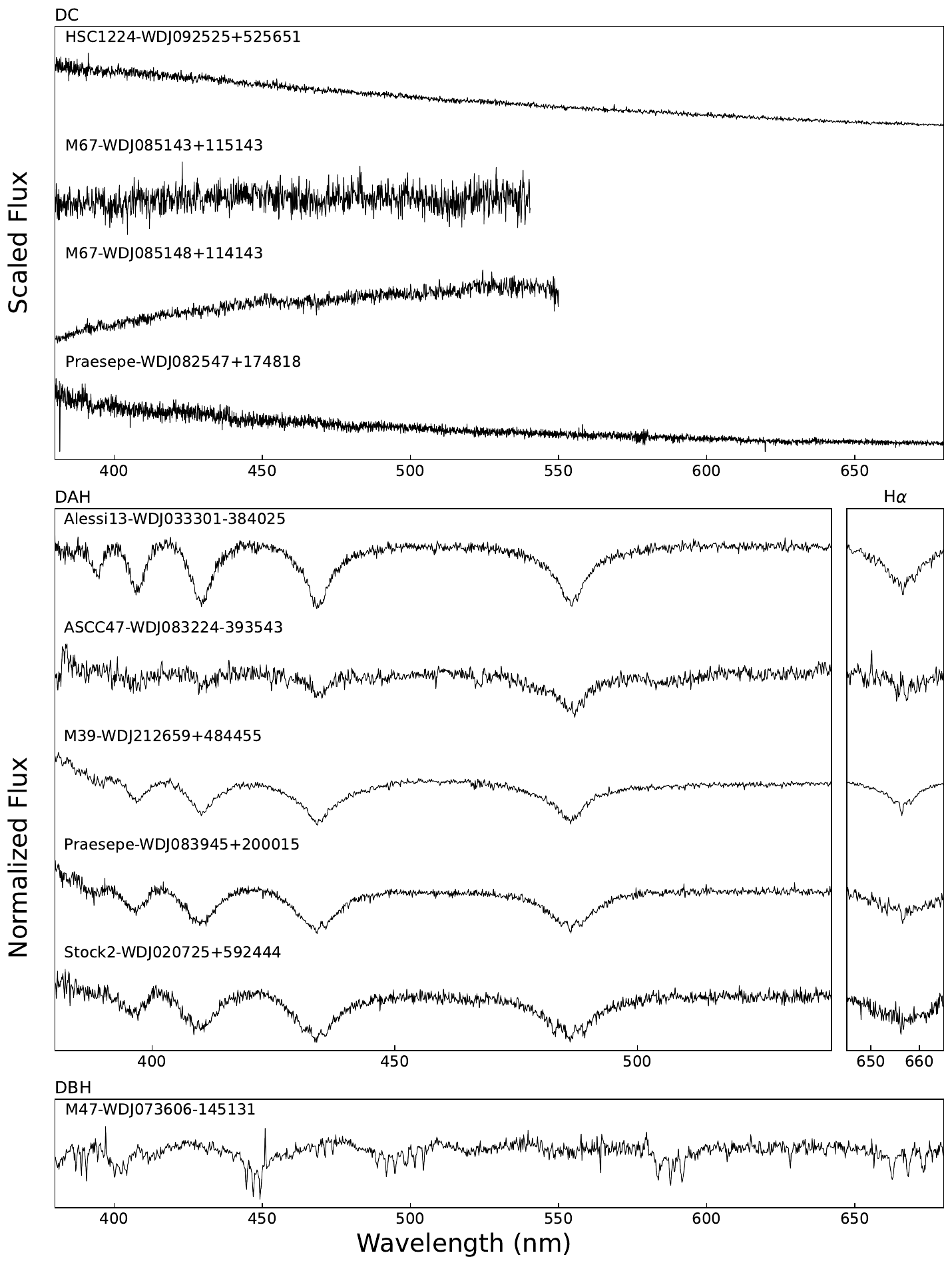}
    \caption{Spectra of the candidate cluster DC and magnetic WDs. Top: DC WDs. Middle: magnetic DAH WDs, with H$\alpha$ shown separately on the right. Bottom: the magnetic DBH WD. The DC spectra are linearly scaled for display without continuum normalization, preserving broad features, while the magnetic spectra are continuum-normalized. Hyades-WDJ041309+185936 is not shown because the spectrum used for its DC classification is not publicly available and could not be obtained for this work; its classification is taken from \citet{1986AJ.....91..139W}.}
    \label{fig:DC_magnetic_spectra}
\end{figure*}

\subsubsection{Photometric Fitting}
\label{subsubsec:photometric_fitting}

For the DC and magnetic WDs, we instead determine atmospheric parameters from available photometry. For the DC WDs, we fit both the pure-H models of \citet{2011ApJ...730..128T} and the mixed H/He models of \citet{2021MNRAS.501.5274C}, adopting $\log(\mathrm{H}/\mathrm{He})=-5$, while for the magnetic WDs we use the corresponding pure-H or pure-He models according to the spectroscopic classification. The synthetic fluxes are scaled using the Gaia parallax, and synthetic photometry is computed with \texttt{pyphot} \citep{fouesneau2025pyphot}. We fit for $T_\textrm{eff}$ and radius using a Levenberg--Marquardt least-squares routine, fixing the reddening to the adopted value for the host cluster.

When available, we use GALEX \citep{2005ApJ...619L...1M} together with Pan-STARRS \citep{2016arXiv161205560C} photometry. If GALEX is unavailable but both Pan-STARRS and SDSS \citep{2000AJ....120.1579Y} are available, we use the Pan-STARRS bands together with the SDSS $u$ band, which improves the temperature constraint for cool WDs \citep{2019ApJ...876...67B}. If Pan-STARRS is unavailable we use SDSS for optical bands. When these bands are unavailable, we substitute suitable optical or near-UV photometry from other surveys or targeted observations, as necessary.

\medskip
\noindent\textit{DC WDs.} The spectra establish the featureless classifications (see the top panel of Fig.~\ref{fig:DC_magnetic_spectra}). HSC1224-WDJ092525+525651 has previously been classified as a DQ by \citet{2019ApJ...885...74C} and \citet{2026ApJ..1006..248K}; however, we retain the DC classification because the higher-S/N SDSS spectrum used here shows no convincing carbon features, including the expected C$_2$ Swan bands. Hyades-WDJ041309+185936 does not have an available spectrum for our analysis, but was classified as a DC WD based on spectra obtained by \citet{1986AJ.....91..139W}. The lack of detectable lines means that spectroscopic line-profile fitting is not possible, and the DC classification itself does not differentiate between H- and He-dominated atmospheres. We therefore fit both H- and He-atmosphere models and adopt the solution whose model spectrum remains consistent with the observed featureless classification.

Under the stated photometric priority scheme, HSC1224-WDJ092525+525651 is fit with Pan-STARRS plus SDSS $u$, Hyades-WDJ041309+185936 with Pan-STARRS plus GALEX NUV, M67-WDJ085148+114143 with SDSS, and Praesepe-WDJ082547+174818 with Pan-STARRS plus GALEX. M67-WDJ085143+115143 has only two usable SDSS bands and no Pan-STARRS or GALEX coverage, so we instead use the SDSS-like $u$, $g$, and $r$ photometry obtained with the Megacam imager on the 6.5~m MMT telescope by \citet{2018ApJ...867...62W}.

\begin{figure}[ht]
\centering
\includegraphics[width=\columnwidth]{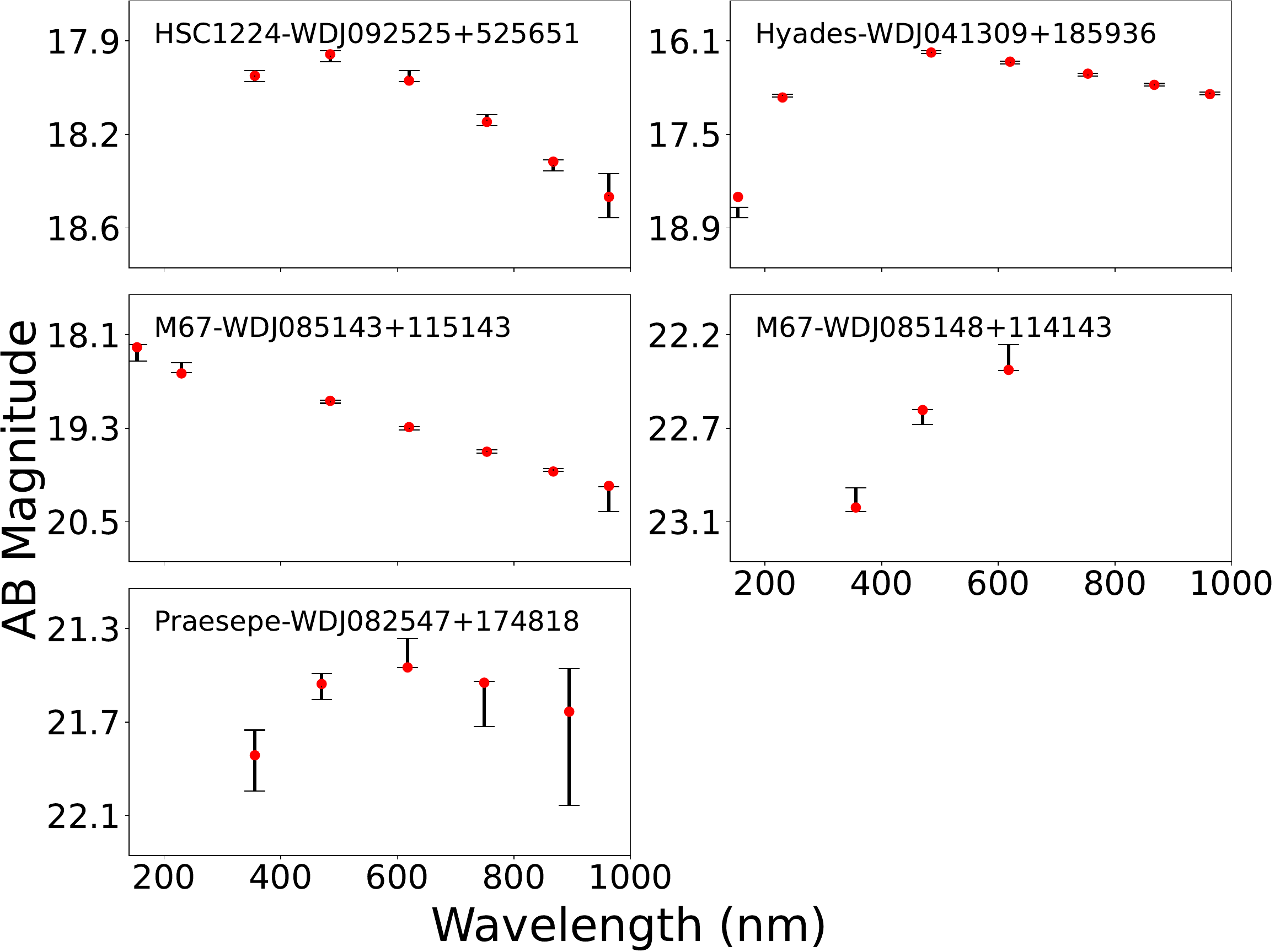}
\caption{Photometric fits of the spectroscopically classified DC WDs, using the available photometry with 1$\sigma$ error bars. Synthetic photometry for the adopted atmospheric composition is shown in red. }
\label{fig:dc_sed_fits}
\end{figure}

For HSC1224-WDJ092525+525651, Hyades-WDJ041309+185936, M67-WDJ085143+115143, and M67-WDJ085148+114143, the H-atmosphere fits predict strong Balmer features that are not present in the observed spectrum, while the He-atmosphere fits predict much weaker or no detectable features. We therefore adopt the He-atmosphere solutions for these objects. Praesepe-WDJ082547+174818 does not yield a self-consistent non-magnetic photometric solution. Both H and He-atmosphere fits imply temperatures at which strong spectral features should be present, in clear conflict with the observed featureless spectrum. We therefore conclude that the WD is highly magnetic, as strong magnetic fields can substantially broaden or suppress the spectral features predicted by non-magnetic atmosphere models. \citet{2026ApJ..1000..216K} classify this source as a DAH; although our SED analysis likewise indicates strong magnetism, we retain the DC spectral classification because the available spectrum does not show convincing Balmer features or Zeeman splitting. Given that most magnetic WDs have DA atmospheres \citep{2025ApJ...990...25M}, we pragmatically adopt the DA solution from the SED fit for the subsequent analysis. The adopted SED fit for each of the five DC WDs is shown in Fig.~\ref{fig:dc_sed_fits}.

\begin{figure}[ht]
\centering
\includegraphics[width=\columnwidth]{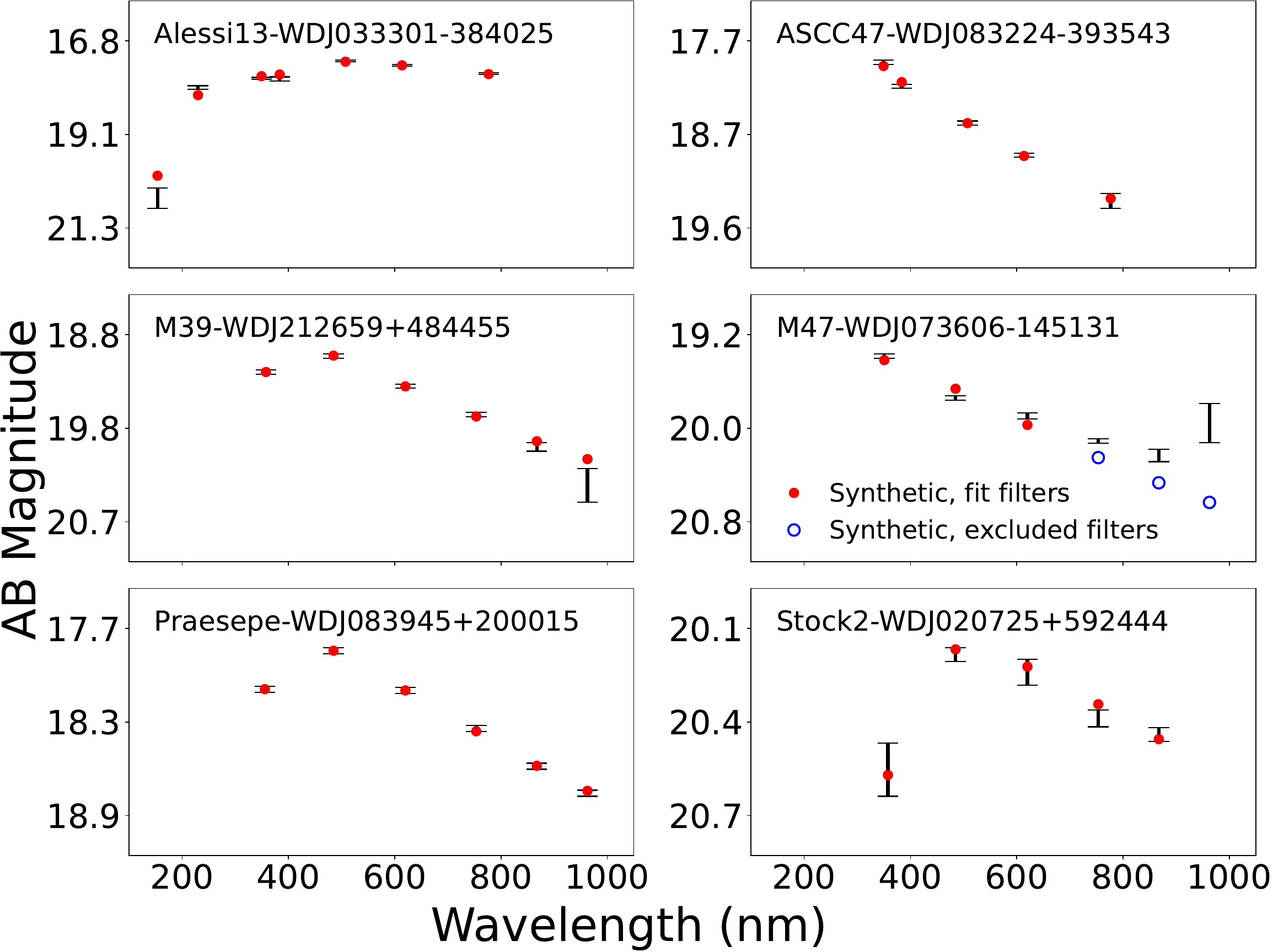}
\caption{Photometric fits for the low-field magnetic WDs, using the available photometry with 1$\sigma$ error bars. Red points show the synthetic photometry from the best-fitting model. For M47-WDJ073606-145131, the photometric points depicted in blue are excluded from the fit because the excess at longer wavelengths is likely due to a late-type companion, as discussed in \citet{2019ApJ...880...75R}.}
\label{fig:magnetic_sed_fits}
\end{figure}

\medskip
\noindent\textit{Magnetic DA/DB WDs.} Spectra for the sample magnetic WDs are shown in the middle and bottom panels of Fig.~\ref{fig:DC_magnetic_spectra}. Magnetic field estimates are based on the observed Zeeman splitting of H$\alpha$ and are summarized in Table~\ref{tab:magnetic_fields}. We determine atmospheric parameters from SED fitting rather than line-profile fitting. This provides reasonable parameter estimates for weakly magnetic WDs because the flux distribution is less affected by the magnetic field than the line profiles, where magnetic broadening is degenerate with surface gravity. The resulting SED fits are shown in Fig.~\ref{fig:magnetic_sed_fits}.

\begin{table}
\tiny
\centering
\caption{Magnetic characteristics for candidate low-field magnetic cluster WDs.}
\label{tab:magnetic_fields}
\begin{tabular}{lccc}
\toprule
Name & Type & $B^\dagger$ & $B$ source \\
& & [MG] & \\
\midrule
Alessi13-WDJ033301-384025 & DAH & 1.1 & this work \\
ASCC47-WDJ083224-393543   & DAH & 1.6 & \citet{2020ApJ...901L..14C} \\
M39-WDJ212659+484455      & DAH & 1.4 & \citet{2020ApJ...901L..14C} \\
M47-WDJ073606-145131      & DBH & 2.5 & \citet{2019ApJ...880...75R} \\
Praesepe-WDJ083945+200015 & DAH & 2.2 & \citet{2013MNRAS.429.2934K} \\
Stock2-WDJ020725+592444   & DAH & 3.0 & this work \\
\bottomrule
\end{tabular}
\begin{minipage}{\linewidth}
\medskip
{\textbf{Notes}. $^\dagger$ Magnetic field strengths are approximate visual estimates from the cited sources or from this work.}
\end{minipage}
\end{table}

Alessi13-WDJ033301-384025 and ASCC47-WDJ083224-393543 lack the preferred optical bands and are instead fit with SkyMapper photometry \citep{2007PASA...24....1K,2024PASA...41...61O}; for Alessi13-WDJ033301-384025, we additionally include GALEX. At the high temperature of ASCC47-WDJ083224-393543, the SED provides only a weak temperature constraint. We therefore use the spectroscopic temperature determined from the magnetic atmosphere models of \citet{2020ApJ...901L..14C}, and combine it with our SED-derived radius to determine the remaining parameters. M39-WDJ212659+484455 is fit with Pan-STARRS plus UVEX $u$ photometry from the INT Galactic Plane Survey \citep[IGAPS;][]{2020A&A...638A..18M}. M47-WDJ073606-145131 is fit with Pan-STARRS plus VPHAS+ $u$ \citep{2014MNRAS.440.2036D}, with the three redder Pan-STARRS bands neglected due to red excess likely from a late-type companion \citep{2019ApJ...880...75R}. Praesepe-WDJ083945+200015 is fit with Pan-STARRS plus SDSS $u$. For Stock2-WDJ020725+592444, we use our own $u$-band photometry from an unpublished UVEX field \citep{2009MNRAS.399..323G}; because Stock~2 has significant differential reddening, we marginalize over the $1\sigma$ reddening range rather than fixing reddening.


\subsection{Cooling Parameters and Progenitor Masses}
\label{subsec:coolingparams}

For the adopted WD parameters, we use the cooling models by \citet{2020ApJ...901...93B} to complete the mapping between atmospheric parameters, radius, mass, and cooling age. For spectroscopic fits, the inputs are $T_\textrm{eff}$ and $\log g$, which give the corresponding mass, radius, and cooling age. For photometric fits, the fitted quantities are $T_\textrm{eff}$ and radius, so the same model grid is used to infer $\log g$, mass, and cooling age. We use the thin-H evolutionary models for the DB/DBA WDs and for DC WDs with adopted He-atmosphere solutions, and the corresponding thick-H envelope models for objects with adopted H-atmosphere solutions.

Where a single-star cluster interpretation is favored, we estimate progenitor masses by subtracting the WD cooling age from the adopted cluster age to obtain the progenitor lifetime. We then use PARSEC isochrones at the adopted cluster metallicity to infer the corresponding initial mass. As in \miller, the progenitor mass is defined as the lowest-mass star that has reached the onset of the asymptotic giant branch. 

Recent population studies increasingly favor multiple formation channels for magnetic WDs, with young, high-mass, high-field WDs associated with merger formation, while older, more average-mass WDs with weaker fields are consistent with single-star formation channels \citep[and references therein]{2025ApJ...990...25M}. Additionally, low-field massive magnetic WDs in young clusters have also been shown to be consistent with single-star evolution \citep{2020ApJ...901L..14C}. We therefore do not exclude magnetic WDs from the progenitor-mass calculation solely on the basis of magnetism. When their WD parameters and cluster ages remain consistent with single-star evolution, we report progenitor masses in the same way as for non-magnetic WDs. We do not report progenitor masses when the cooling age and cluster age do not yield a meaningful single-star progenitor lifetime, or when the WD properties instead point toward binary evolution or a merger origin.

We do not report progenitor masses for the two low-mass M67 WDs, M67-WDJ085143+115143 and M67-WDJ085148+114143, because these WDs are far too low mass to have formed through single-star evolution within a Hubble time. We likewise omit a progenitor mass for M67-WDJ085029+115249: although DB/DBA atmospheric parameters are more model dependent than those of DAs, its unusually high inferred mass compared with other M67 WDs, which have nearly identical progenitor masses, is suggestive of a binary evolutionary origin. We also omit a progenitor mass for Praesepe-WDJ082547+174818 because its highly magnetic nature makes the $T_\textrm{eff}$ and mass inferred from the adopted non-magnetic atmospheric model uncertain, which in turn makes the cooling age unreliable. Hyades-WDJ041309+185936 has a cooling age slightly exceeding its cluster age, but we retain it given the greater model dependence of the atmospheric and cooling parameters for DC WDs. In contrast, the cooling age of Alessi13-WDJ033301-384025 substantially exceeds the age of Alessi~13, ruling out cluster membership and leaving a final sample of 18 atypical WDs.


\section{The Non-DA Population}
\label{sec:nonDAs}

Of the 18 atypical cluster WDs in our final sample, 14 are non-DAs: five DBs, three DBAs, four non-magnetic DCs, one magnetic DBH, and one spectroscopically featureless WD that we interpret as highly magnetic. Nine of these yield viable single-star progenitor-mass solutions, comprising four DBs, all three DBAs, one DC, and the DBH. Comparable cluster DA samples generally exclude WDs whose properties are inconsistent with single-star evolution, so including these five objects would compare a broader non-DA population against a more restrictive DA population. We therefore restrict the comparisons below to the nine non-DAs with viable single-star solutions and compare them with a DA sample assembled from the 122 DA WDs in \miller, one additional DA from \citet{2026ApJ..1007..183M}, and nine from the archival spectroscopic census of \citet{Miller2026Surveys}. Data for the 123 previously published DAs are available at wdifmr.org\footnote{\url{https://wdifmr.org}} and the nine DAs analyzed in \citet{Miller2026Surveys} will be added upon acceptance of that manuscript. We first compare the non-DA incidence in clusters and the field, then examine their locations in the IFMR, and finally compare the spatial and kinematic distributions of the non-DA and DA cluster populations.


\subsection{Comparison with Field WDs}
\label{subsec:nonDA_field}

\begin{figure}[ht]
    \centering
    \includegraphics[width=\columnwidth]{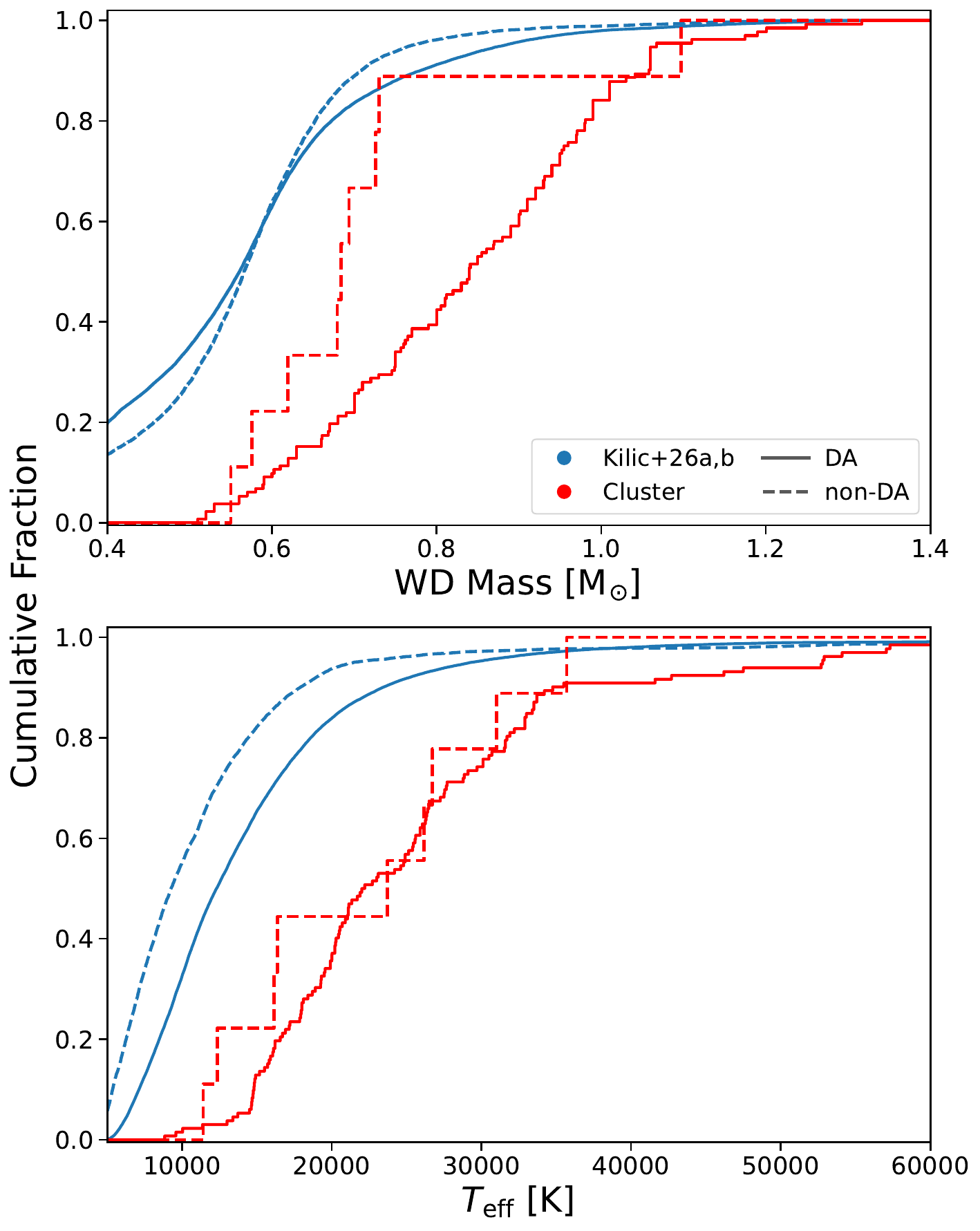}
    \caption{Cumulative WD mass (top) and effective temperature (bottom) distributions for the combined DESI field sample of \citet{2026ApJ..1006..248K,2026ApJ..1000..216K} and the cluster sample. Black and red curves show the respective samples, while solid and dashed lines indicate the DA and non-DA populations within those samples, respectively. The cluster sample is restricted to WDs with viable single-star progenitor mass solutions.}
    \label{fig:DA_nonDA_field_dist}
\end{figure}

We compare the cluster sample with the combined DESI WD samples of \citet{2026ApJ..1006..248K,2026ApJ..1000..216K}, which provide spectroscopic classifications for a large magnitude-limited field population spanning a similar temperature range to the cluster WDs. We use the DESI spectral classifications to divide the field sample into DAs and non-DAs, retaining magnetic and other subtypes within the corresponding atmospheric group. Binary systems are excluded, as are subdwarfs. We adopt the photometrically determined atmospheric parameters and masses, as the DESI spectroscopic parameters show systematic offsets, particularly in the inferred masses of DA WDs \citep{2026ApJ..1000..216K, Miller2026Surveys}. Fig.~\ref{fig:DA_nonDA_field_dist} compares the mass and effective-temperature distributions of the DA and non-DA populations.

The DESI field sample shows a strong temperature dependence, with non-DAs being cooler on average than DAs, while the mass distributions of the two spectral groups are similar. The cluster population is markedly different. Cluster non-DAs are concentrated at low and intermediate masses, while the high-mass population is almost entirely DA. The cluster sample is also substantially hotter than the field samples, as few clusters are both nearby and old enough to contain readily detectable cool WDs. This raises the possibility that the apparent high-mass non-DA deficit simply reflects the temperature dependence of the non-DA fraction.

\begin{figure}[ht]
    \centering
    \includegraphics[width=\columnwidth]{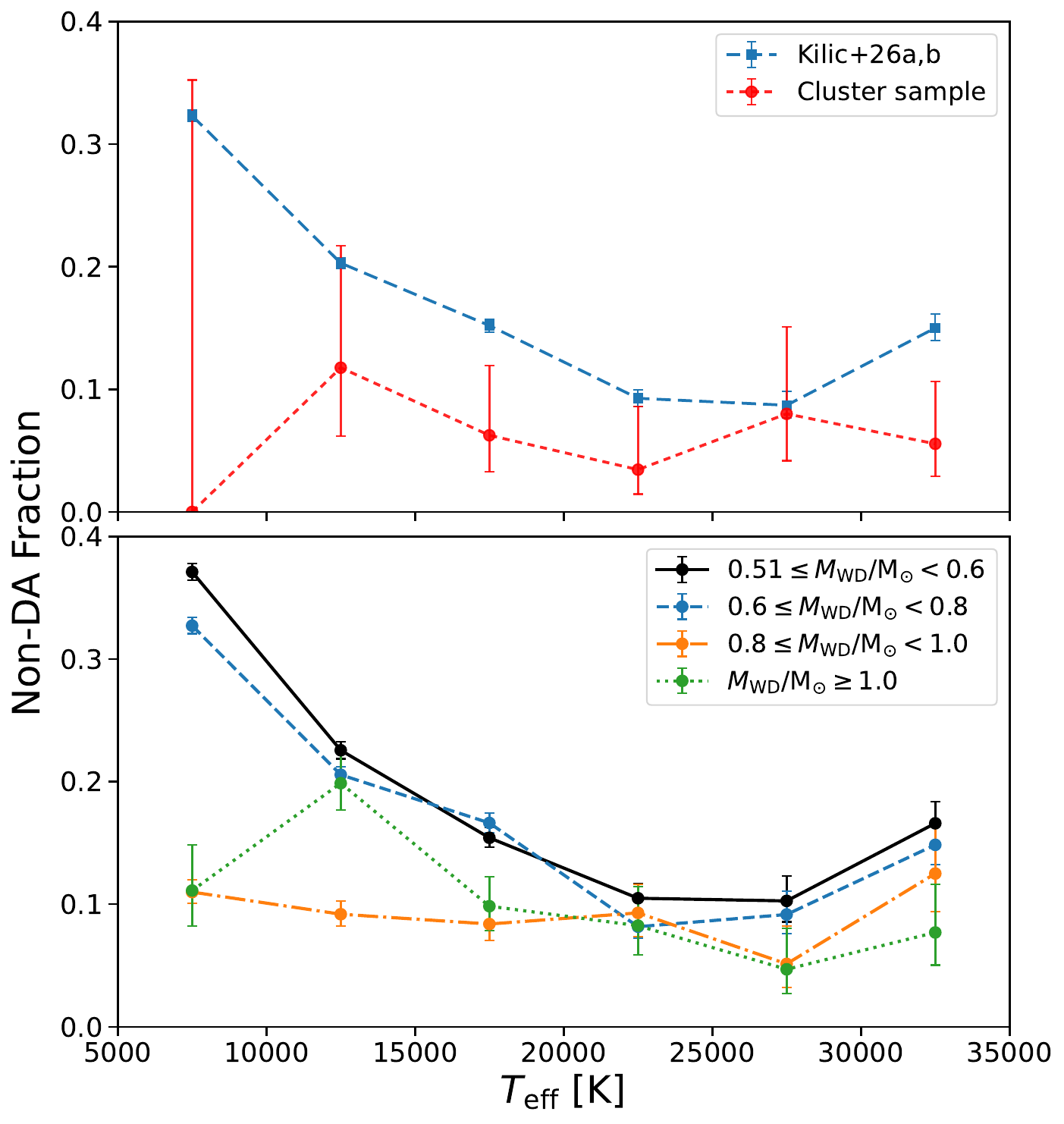}
    \caption{Non-DA fractions across effective temperature bins. Top: fractions for the field (blue) and cluster samples (red). Bottom: fractions for the field WD sample, divided into four WD mass ranges. Error bars show 68\% Bayesian credible intervals.}
    \label{fig:nonDA_frac_combined}
\end{figure}

We test this possibility in Fig.~\ref{fig:nonDA_frac_combined}. We restrict the comparison to $M_{\rm WD}\geq0.51\msun$ to avoid contamination from unresolved binaries and calculate non-DA fractions in 5000~K bins between 10,000 and 30,000~K, with cooler and hotter WDs grouped into single bins at the lower and upper ends, respectively. The top panel shows the resulting fractions with 68\% Bayesian credible intervals from a Jeffreys binomial prior. The cluster non-DA fraction lies below that of the DESI field sample in every temperature bin, although the uncertainties overlap in several bins. To examine whether this difference extends to massive WDs, the lower panel divides the field sample into four mass ranges. The non-DA fraction remains substantial at high mass over much of this temperature range, so the hotter temperature distribution of cluster WDs cannot by itself explain the lack of massive non-DAs.

\begin{figure*}[ht]
    \centering
    \includegraphics[width=\textwidth]{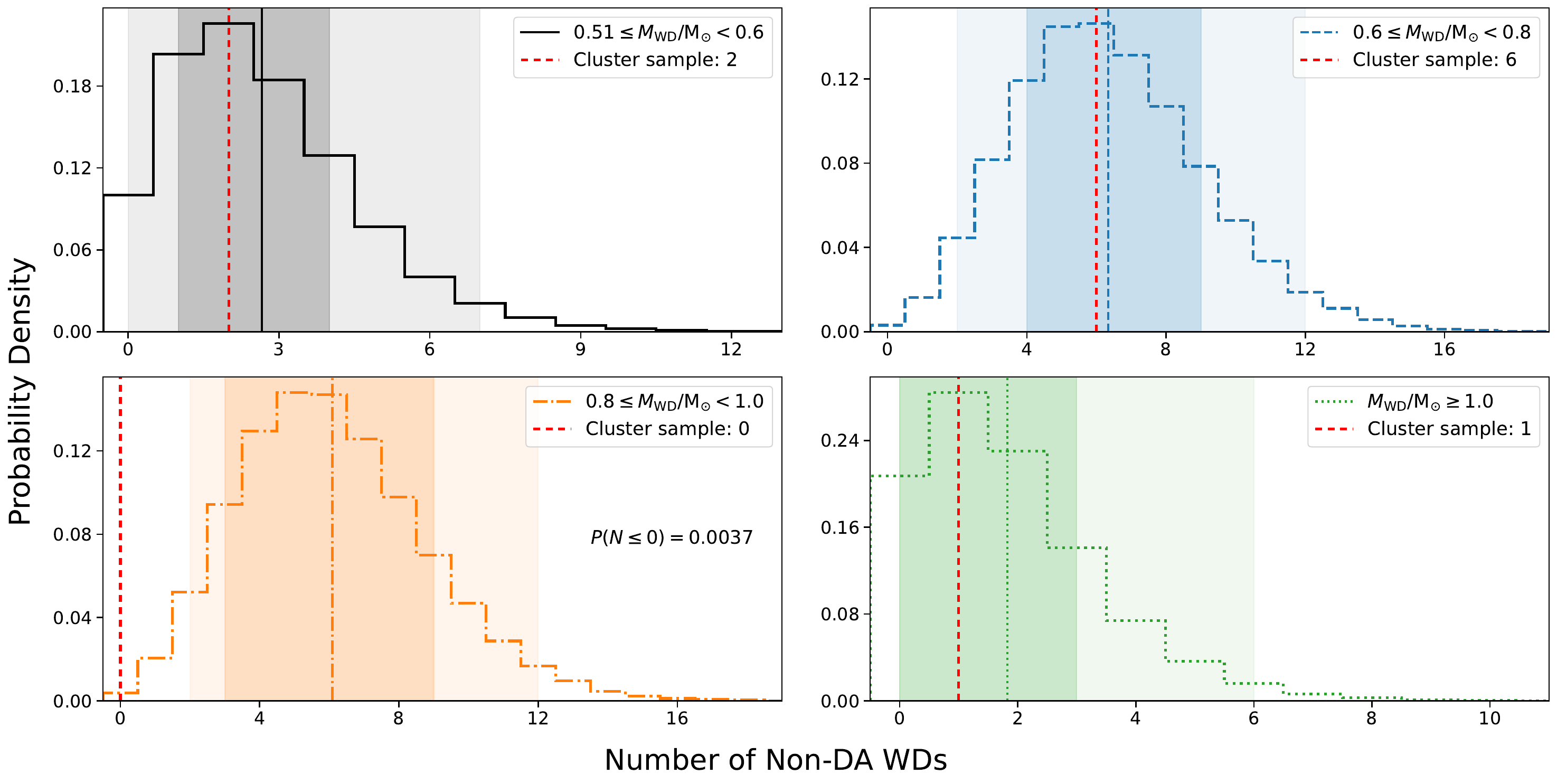}
    \caption{Posterior predictive distributions for the expected numbers of non-DA cluster WDs in four WD mass ranges, inferred from the temperature-dependent fractions of the DESI field sample. Shaded regions mark the 68\% and 95\% intervals, vertical lines indicate the predicted means, and red dashed lines show the observed cluster counts. In the bottom left panel, $P$ gives the probability of obtaining a count less than or equal to the observed value.}
    \label{fig:nonDA_expected_bymass}
\end{figure*}

We quantify this mass dependence by using the field non-DA/DA ratio in each mass and temperature bin to predict the number of non-DAs associated with the observed cluster DA population. Repeating this calculation 100,000 times gives the distributions in Fig.~\ref{fig:nonDA_expected_bymass}. The two lower-mass bins, along with the highest-mass bin, are each consistent with the field expectation. In contrast, at $0.8\leq M_{\rm WD}/\msun<1.0$, we predict $6.0^{+0.8}_{-0.7}$ non-DAs but observe none, with a posterior-predictive probability of $P(N=0)=3.7\times10^{-3}$. The cluster non-DA deficit is therefore not a general reduction in the non-DA fraction, but a strongly mass-dependent effect concentrated between $0.8$ and $1.0\msun$.

The missing high-mass population is unlikely to consist primarily of ordinary DBs. Field samples also contain very few DBs above $0.8\msun$, with more massive DBs generally magnetic or rapidly rotating \citep{2019ApJ...882..106G,2024ApJ...974...12J,2025ApJ...979..157K}. Ordinary cool DCs are also unlikely to account for the deficit, as only four DAs used to predict the cluster non-DA counts fall below $12{,}000$~K, the approximate temperature below which non-magnetic DCs are typically found. This points to DQs and peculiar magnetic WDs as important contributors to the high-mass field non-DA population. In a spectroscopic follow-up of 112 Gaia-selected massive DB candidates, \citet{2026A&A...711A..70G} confirmed only five as normal DB/DBA WDs; the classifications instead included 51 magnetic WDs, 24 DCs, and 27 warm or hot DQs. \citet{2024ApJ...974...12J} similarly find 20 DQs and 66 magnetic WDs among 204 WDs with $M_{\rm WD}>0.9\msun$ within 100~pc, although 50 of those magnetic WDs are DAs. We examine magnetism in the cluster sample in more detail in Section~\ref{sec:magnetic_wds}.

We next consider whether selection or measurement biases could explain the deficit. Cluster studies often use consistency with single-star evolution as additional evidence for membership, so merger remnants and other peculiar WDs may be rejected as candidates or receive less targeted spectroscopic follow-up. However, relaxing the single-star consistency requirement does not remove the high-mass non-DA deficit: the broader non-DA sample still lacks the required population between $0.8$--$1.0\msun$, while any corresponding population of DAs requiring binary evolution that is absent from the comparison sample would increase, rather than decrease, the number of non-DAs expected from the field ratio. A separate concern is the differing mass precision of the two samples, as the DESI field masses have somewhat larger uncertainties than those of the cluster WDs, which could increase the scattering between adjacent mass bins. As a conservative test of the differing mass precision, we degraded the cluster mass uncertainties to match the empirical field distribution within each temperature bin and repeated the comparison; the deficit reduced but remained highly significant. These effects may reduce the inferred significance of the deficit, but the complete absence of non-DAs from $0.8$--$1.0\msun$ remains difficult to attribute to selection and measurement uncertainties alone.

The remaining difference may instead reflect the different evolutionary timescales sampled by clusters and the field. The DA comparison sample is strongly weighted towards young clusters, with only eight host clusters older than 750~Myr. This is particularly relevant for massive DQs, whose high masses and unusual kinematics have been linked to merger evolution. Their standard cooling ages are typically of order 1~Gyr, while their kinematics indicate substantially older ages and multi-Gyr cooling delays \citep{2020ApJ...891..160C}. Most clusters with known WDs may therefore be too young to contain a substantial population of these objects, while those old enough may have lost them to the field through open-cluster disruption \citep[e.g.,][]{2005A&A...441..117L}.

Overall, the cluster non-DA deficit is concentrated at high WD mass and is strongest between $0.8$ and $1.0\msun$. Selection and measurement error effects may reduce its statistical significance, but appear unlikely to explain it entirely. Instead, the deficit may reflect the absence of massive DQs and other peculiar non-DAs that arise or become observable on evolutionary timescales poorly sampled by the predominantly young population of clusters with known WDs. The young ages of most WD-hosting clusters, combined with the disruption of older clusters, may therefore explain why these populations are present in the field but largely absent from clusters.


\subsection{The IFMR for Non-DA WDs}
\label{subsec:nonaDA_ifmr}

\begin{figure*}[ht]
    \centering
    \includegraphics[width=\textwidth]{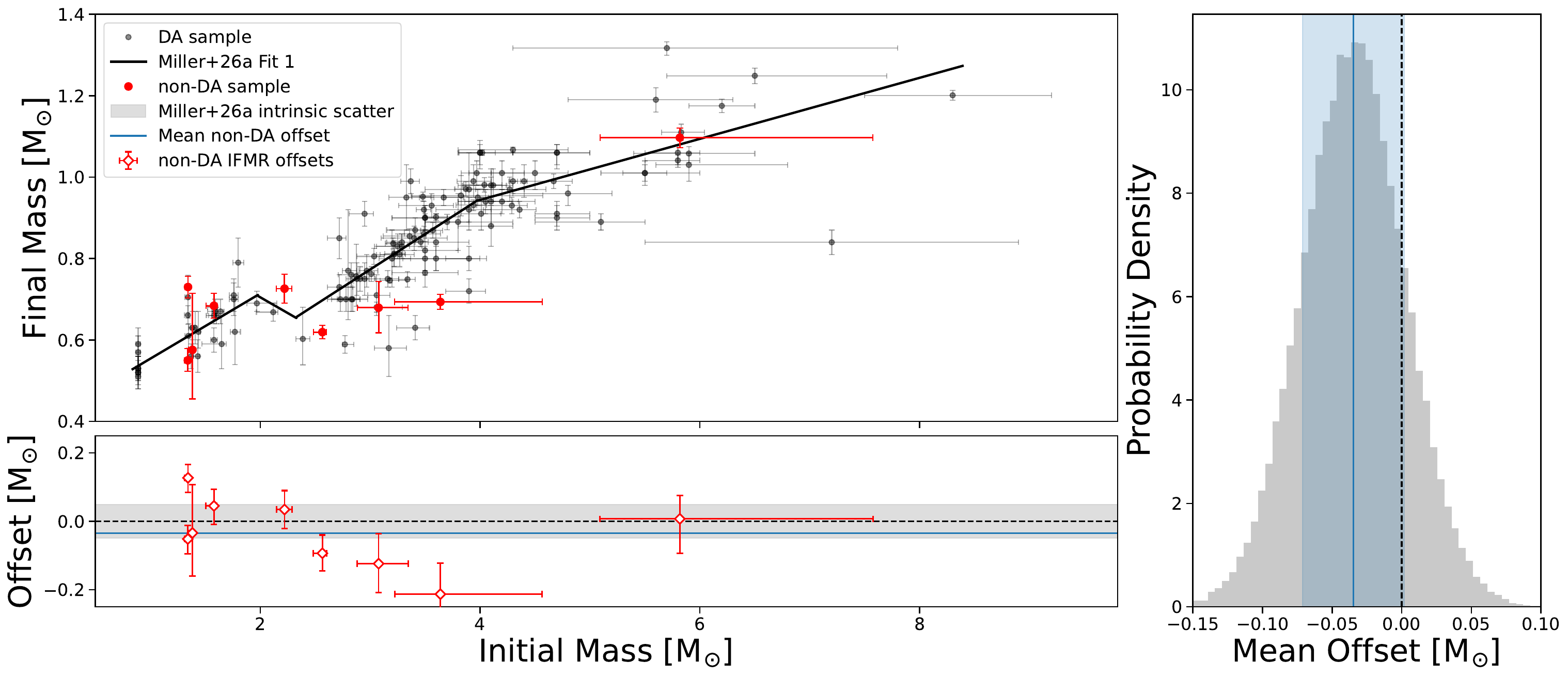}
    \caption{Top left: IFMR for the DA and non-DA cluster WD samples. The DA comparison sample is shown in gray, with the best-fit relation from \miller shown as the black line. The non-DA WDs for which we derive progenitor masses are shown in red. Bottom left: residuals of the non-DA sample, defined as $\Delta M_f = M_f-M_{f,\mathrm{IFMR}}$, with uncertainties from Monte Carlo sampling of the WD masses, progenitor masses, and published IFMR parameters. The gray region marks the intrinsic scatter of the \miller DA relation, and the blue line marks the median inferred mean offset. Right: Monte Carlo distribution of the mean non-DA offset, including the adopted DA intrinsic scatter. The blue line marks the median inferred mean offset, while the blue band gives the central 68\% interval.}
    \label{fig:nonDA_ifmr}
\end{figure*}

The sample of nine non-DA WDs with single-star progenitor-mass solutions is too small to examine the form of a non-DA IFMR. That said, we can test whether the sample is statistically consistent with the trend derived from cluster DA IFMR samples. The left panel of Fig.~\ref{fig:nonDA_ifmr} shows the non-DA sample alongside the DA comparison sample, with the Full Sample Fit~1 relation from \miller overlaid. Visually, there is no obvious systematic displacement of the non-DA WD sample from their best-fit relation.

To quantify this comparison, we calculate the residual $\Delta M_f=M_{f,\mathrm{non-DA}}-M_{f,\mathrm{IFMR}}$ for each non-DA WD using Monte Carlo sampling. For each WD, we draw progenitor and WD masses from split-normal distributions of the best-fit values and their asymmetric uncertainties. We similarly draw from the parameters of the \miller Full Sample Fit~1 from its published uncertainties. A single realization of this relation is applied to all nine non-DA WDs in each Monte Carlo iteration. We then calculate the residual of each WD from the IFMR, and from that, the mean residual of the non-DA sample. This procedure is repeated for 100,000 realizations.

We additionally account for the intrinsic scatter measured about the DA IFMR, $\sigma_{\rm int}=0.049\msun$, in \miller. For each Monte Carlo realization, we add a Gaussian deviation with $\sigma=0.049\msun$ to each residual before calculating the mean offset. The resulting residuals and distribution of the mean offset are shown in the bottom-left and right panels of Fig.~\ref{fig:nonDA_ifmr}. We find a mean offset of $\Delta M_f=-0.035\pm0.037\msun$, with approximately $17\%$ of the Monte Carlo realizations yielding a positive mean offset. The non-DA sample shows a mild tendency toward lower final masses at fixed progenitor mass, but the inferred offset is consistent with zero. We find no significant evidence that the non-DA WDs are systematically displaced from the DA cluster WD IFMR. A greatly increased sample of non-DA cluster WDs with well-constrained progenitor masses will be required to test for more subtle differences, if any, between the two populations.


\subsection{Spatial and Kinematic Distributions}
\label{subsec:nonDA_spatial_kinematic}

To test for differences in the spatial and kinematic distributions of cluster DA and non-DA WDs, we compare the two populations with each other and with the broader populations of their host clusters. We restrict this analysis to WDs with Gaia DR3 counterparts whose host clusters are included in the \citet{2024A&A...686A..42H} catalog, which excludes ASCC~47 and AB~Doradus. We do not impose a membership probability cut, as the probabilities assigned by the clustering algorithm depend strongly on cluster-centric distance and such a cut would preferentially remove members at larger radii. We further exclude six DAs identified through traceback analyses: one Hyades WD \citep{2023ApJ...956L..41M}, three Alpha~Persei WDs \citep{2022ApJ...926L..24M}, and two Pleiades WDs \citep{2022ApJ...926..132H}. The host-cluster member samples do not include a corresponding population identified through similar methods, so including these WDs would bias the comparison. The remaining sample contains 84 DAs and eight non-DAs.

For each WD and star listed as a member of its host cluster in the \citet{2024A&A...686A..42H} catalog, we calculate the angular separation from the adopted cluster center and convert it to a projected physical distance, $R_{\rm proj}=d_{\rm cl}\theta$, where $d_{\rm cl}$ is the median cluster distance. We examine $R_{\rm proj}$ directly and also normalized by the Jacobi radius, $r_{\rm J}$, an approximate dynamical boundary, and the half-number radius, $r_{50}$, which traces the observed member distribution, allowing clusters of different physical sizes to be compared on a common scale. We also calculate the enclosed-member fraction, $f_{\rm enc}$, defined as the fraction of members in the cluster with smaller projected radii than the WD.

\begin{figure*}[ht]
    \centering
    \includegraphics[width=\textwidth]{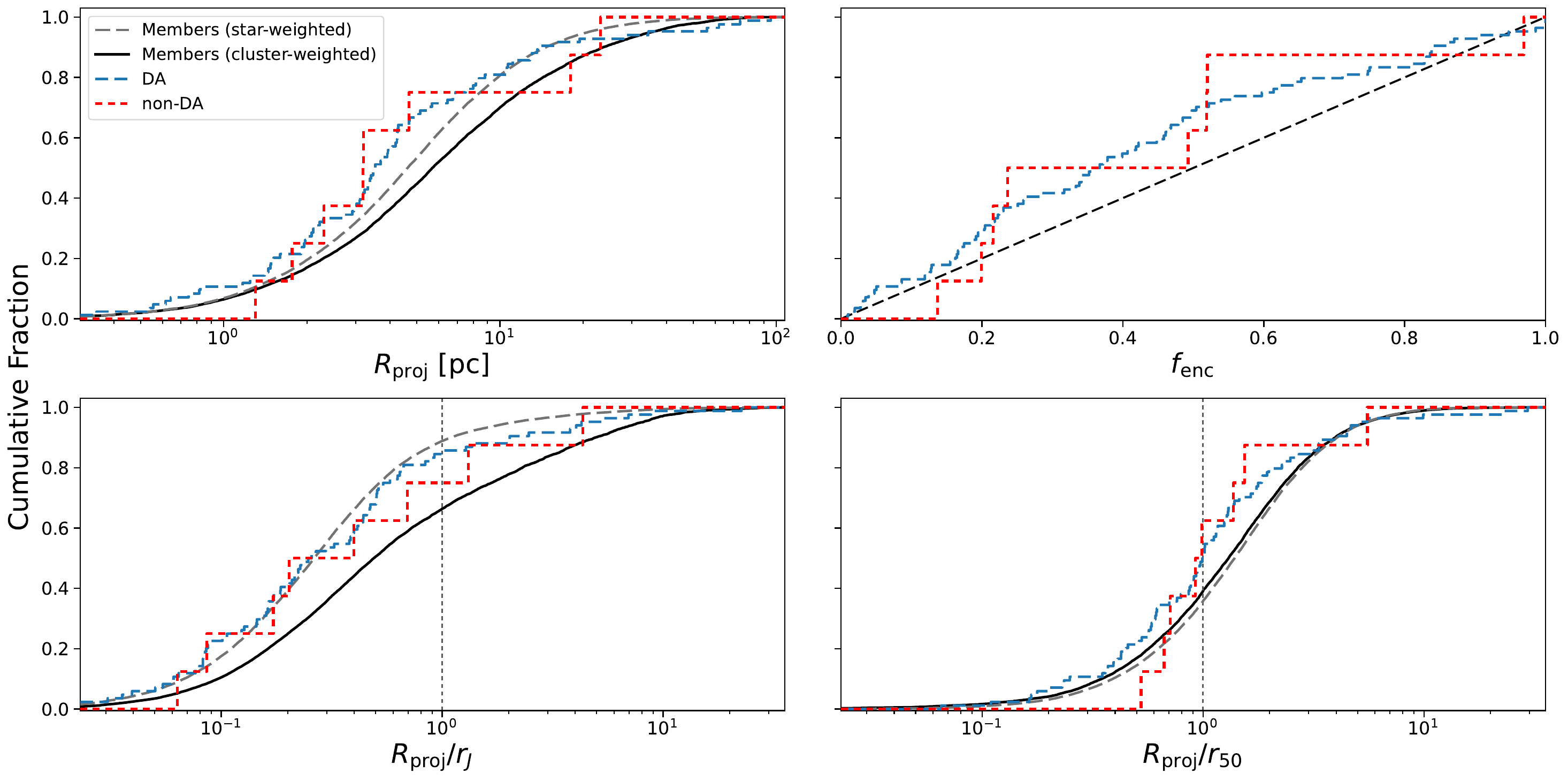}
    \caption{Cumulative spatial distributions of the DA and non-DA cluster WDs compared with the \citet{2024A&A...686A..42H} member populations. Gray dashed and black solid curves show the star- and cluster-weighted member distributions, respectively, while blue and red dashed curves show the DA and non-DA WDs. The diagonal line in the $f_{\rm enc}$ panel shows the uniform distribution expected if the WDs follow the radial distributions of their host-cluster members.}
    \label{fig:nonDA_spatial_dist}
\end{figure*}

We compare the WD radial distributions with cumulative distributions of catalog members in Fig.~\ref{fig:nonDA_spatial_dist}. For $R_{\rm proj}$ and the normalized radii, we show both a star-weighted distribution, in which each member contributes equally and richer clusters therefore contribute more, and a cluster-weighted distribution, in which each member is weighted by $1/N$, where $N$ is the number of catalog members in its host cluster, so that every cluster contributes equally. WDs following the same radial distribution as their host-cluster members would have a uniform $f_{\rm enc}$ distribution.

The DA and non-DA spatial distributions are very similar, with median enclosed-member fractions of 0.362 and 0.365, respectively. A two-sample Kolmogorov--Smirnov (KS) test of the $f_{\rm enc}$ distributions finds no evidence that they differ ($D=0.179$, $p=0.946$). Both samples are shifted toward smaller radii relative to their host-cluster members, most clearly in $R_{\rm proj}/r_{50}$ and $f_{\rm enc}$. A one-sample KS test against the uniform $f_{\rm enc}$ distribution shows that the DAs are significantly more centrally concentrated ($D=0.198$, $p=2.3\times10^{-3}$). The eight non-DAs show a broadly similar tendency, although the difference is not significant ($D=0.355$, $p=0.208$) given the small sample size.

The star- and cluster-weighted member distributions are nearly identical when scaled by $r_{50}$, since this radius is defined from the observed member distribution of each cluster. A larger difference is seen for $R_{\rm proj}/r_{\rm J}$, where the star-weighted distribution is shifted toward smaller radii. This indicates that the richer clusters, which dominate the star-weighted distribution, tend to have their members concentrated within a smaller fraction of their Jacobi radii. The similarity of the WD distributions under both normalizations suggests that their apparent central concentration is not driven solely by differences in cluster size or richness. The central concentration of the WDs may partly reflect mass segregation of their initially more massive progenitors. However, the stricter criteria generally used to identify cluster WDs for spectroscopic follow-up could also favor sources closer to their cluster centers. 

We next compare the kinematics of the DA and non-DA cluster WDs. We use the same sample selection as for the spatial analysis, with the additional requirement of Gaia DR3 proper motions, leaving 81 DAs and seven non-DAs. For each WD and catalog member, we predict the proper motion expected from the bulk motion of its host cluster at the source position, following the convergent-point method described in \citet{2009A&A...497..209V} and including the contribution from the cluster radial velocity. We then calculate the residual proper motion, $\Delta\mu$, relative to this predicted motion and convert it to a transverse velocity using
$\Delta v_{\perp}=4.74047\,d_{\rm cl}\Delta\mu$, with $d_{\rm cl}$ in kpc and $\Delta\mu$ in mas~yr$^{-1}$.

\begin{figure}[ht]
    \centering
    \includegraphics[width=\columnwidth]{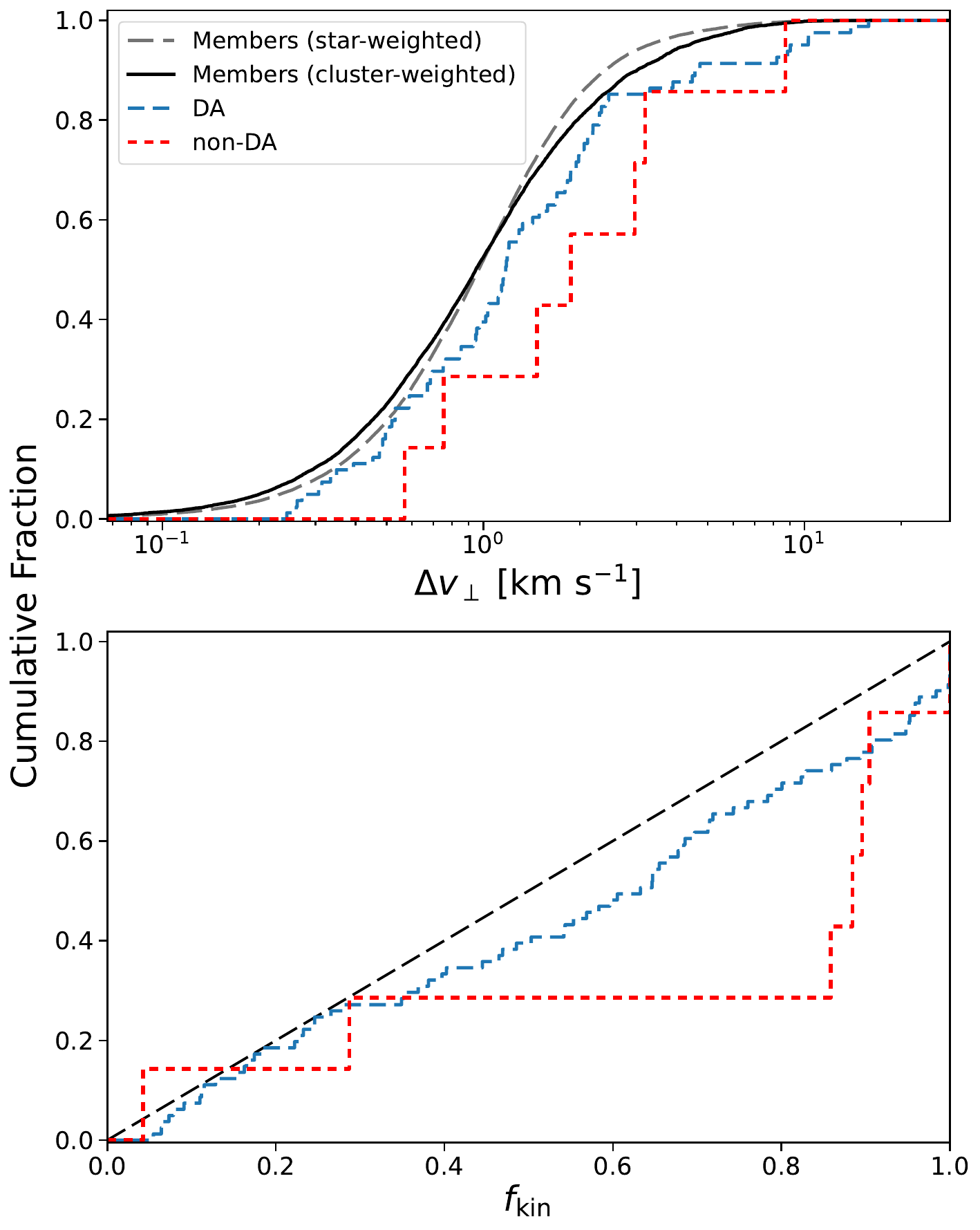}
    \caption{As in Fig.~\ref{fig:nonDA_spatial_dist}, but for the tangential kinematic distributions. $\Delta v_{\perp}$ is the transverse velocity residual relative to the motion predicted from each host cluster.}
    \label{fig:nonDA_kinematics}
\end{figure}

Fig.~\ref{fig:nonDA_kinematics} compares the WD distributions with the cluster members, again using star- and cluster-weighted reference distributions. We also calculate the kinematic percentile, $f_{\rm kin}$, defined as the fraction of members in the same cluster with smaller $\Delta v_{\perp}$. WDs following the same kinematic distribution as the members of their host clusters would have an approximately uniform $f_{\rm kin}$ distribution.

The DAs have median $\Delta v_{\perp}=1.17$~km~s$^{-1}$ and $f_{\rm kin}=0.63$, compared with 1.87~km~s$^{-1}$ and 0.88 for the non-DAs. The non-DAs show larger residual velocities, but a two-sample KS test of the $f_{\rm kin}$ distributions finds no significant difference ($D=0.455$, $p=0.100$). The DA distribution is also consistent with the cluster members ($D=0.140$, $p=0.075$), while the seven non-DAs differ significantly ($D=0.573$, $p=0.0106$).

We also test whether the larger residuals of the non-DAs are inflated by their larger astrometric uncertainties. To do this, we repeat the comparison using residuals normalized by the full Gaia proper-motion covariance and the uncertainties in the cluster mean proper motions. The DA and non-DA distributions remain consistent ($D=0.349$, $p=0.332$), while the difference between the non-DAs and cluster members is no longer significant ($D=0.286$, $p=0.523$). This suggests that the larger raw residuals of the non-DAs are inflated by their astrometric uncertainties, and we find no robust evidence for a kinematic difference between the DA and non-DA cluster WD populations.


\section{The Magnetic WD Population}
\label{sec:magnetic_wds}

Following the exclusion of Alessi13-WDJ033301-384025, the cluster sample contains six magnetic WDs: four DAHs, one DBH, and one spectroscopically featureless WD which we interpret as highly magnetic. We use this sample to compare the magnetic and non-magnetic cluster WD populations with their field counterparts and to examine the implications for magnetic-WD formation channels. We discuss the magnetic DC WD, Praesepe-WDJ082547+174818, in greater detail in Section~\ref{subsec:praesepe_dc_wd} because of a combination of unusual observational properties.

\begin{figure}[ht]
    \centering
    \includegraphics[width=\columnwidth]{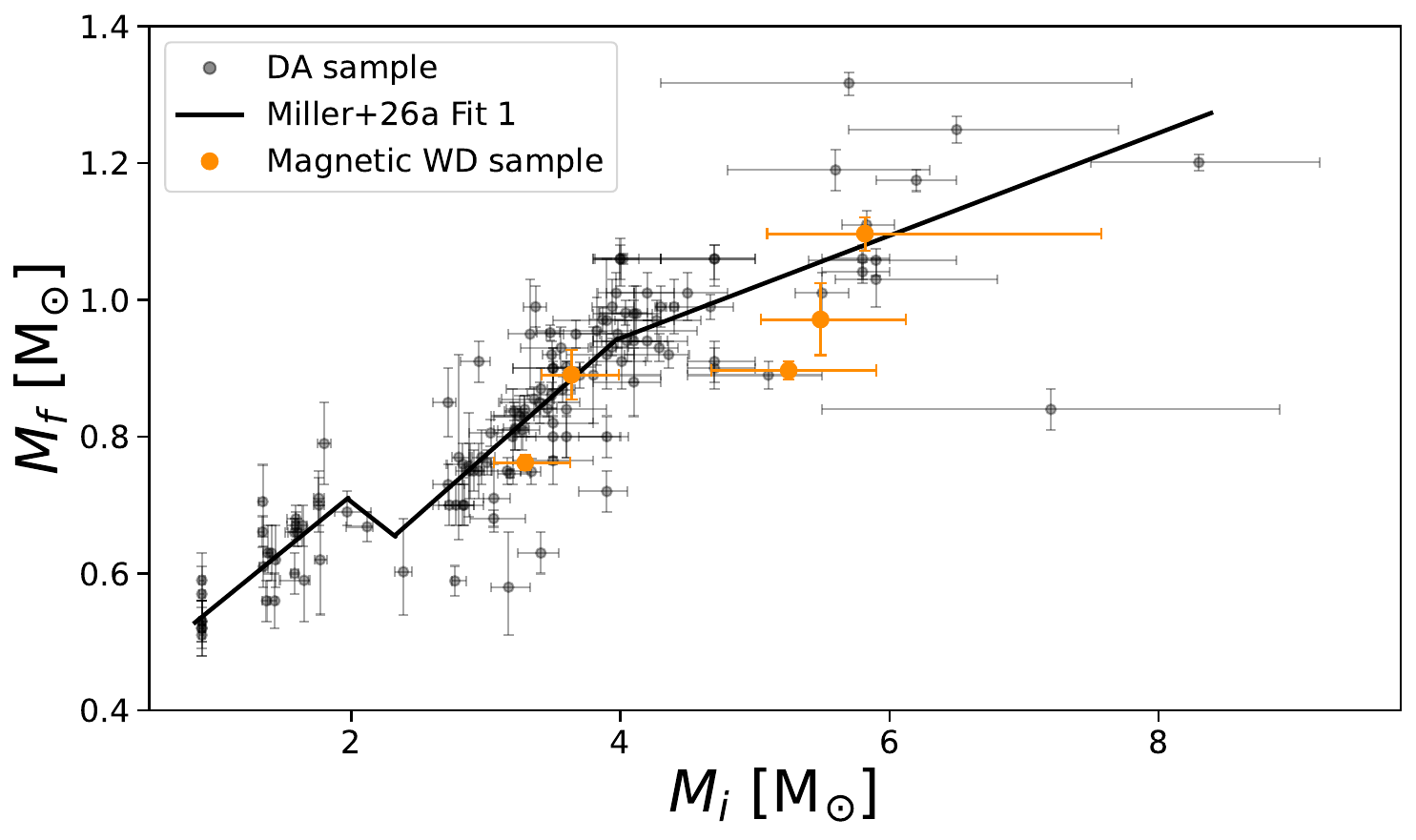}
    \caption{Initial--final mass relation for the magnetic cluster WDs with viable single-star progenitor-mass solutions. Gray points show the non-magnetic DA comparison sample, with the adopted IFMR fit from \miller in black, while orange points show the magnetic WDs.}
    \label{fig:magnetic_ifmr}
\end{figure}

Five of the six magnetic WDs yield viable single-star progenitor-mass solutions and are shown alongside the DA comparison sample in Fig.~\ref{fig:magnetic_ifmr}. Their locations broadly follow the DA IFMR. Three of the five magnetic WDs were already included in the \miller IFMR sample because of their consistency with single-star evolution.

\subsection{Magnetic Incidence}
\label{subsec:magnetic_incidence}

We construct a field comparison sample in the 100~pc SDSS footprint by combining the magnetic WDs and their parameters from \citet{2025ApJ...990...25M} with the corresponding non-magnetic sample from \citet{2025ApJ...979..157K}, which analyzes the same parent population. This gives a volume-limited field sample with magnetic sensitivity reasonably similar to the cluster sample. \citet{2025ApJ...990...25M} identify magnetism primarily through Zeeman splitting, with additional sensitivity to very high-field WDs whose lines are suppressed at temperatures where H or He features would otherwise be expected. Their main incompleteness is to weak fields ($B<1$~MG) and cool high-field DCs. Weak fields would likewise generally escape detection in the cluster spectra and would require spectropolarimetry, while most clusters are too young to have formed cool high-field DCs.

For this comparison, we focus on DA and DB subtypes and their magnetic counterparts, and exclude binary classifications. We also exclude DC WDs because magnetic fields are difficult to identify reliably in cool, featureless spectra, while at higher temperatures strong fields make accurate atmospheric classification difficult. We also exclude DQ and DZ WDs, since neither is present in the cluster sample. These cuts remove the magnetic DC WD Praesepe-WDJ082547+174818, leaving five magnetic cluster WDs. The non-magnetic comparison sample contains the eight DB/DBA WDs identified here and 129 DAs from the DA comparison sample define above, after removing its three magnetic DAs. The corresponding field sample contains 151 magnetic and 1,995 non-magnetic WDs. The resulting magnetic fractions are 3.5\% in clusters and 7.0\% in the field.

\begin{figure}[ht]
    \centering
    \includegraphics[width=\columnwidth]{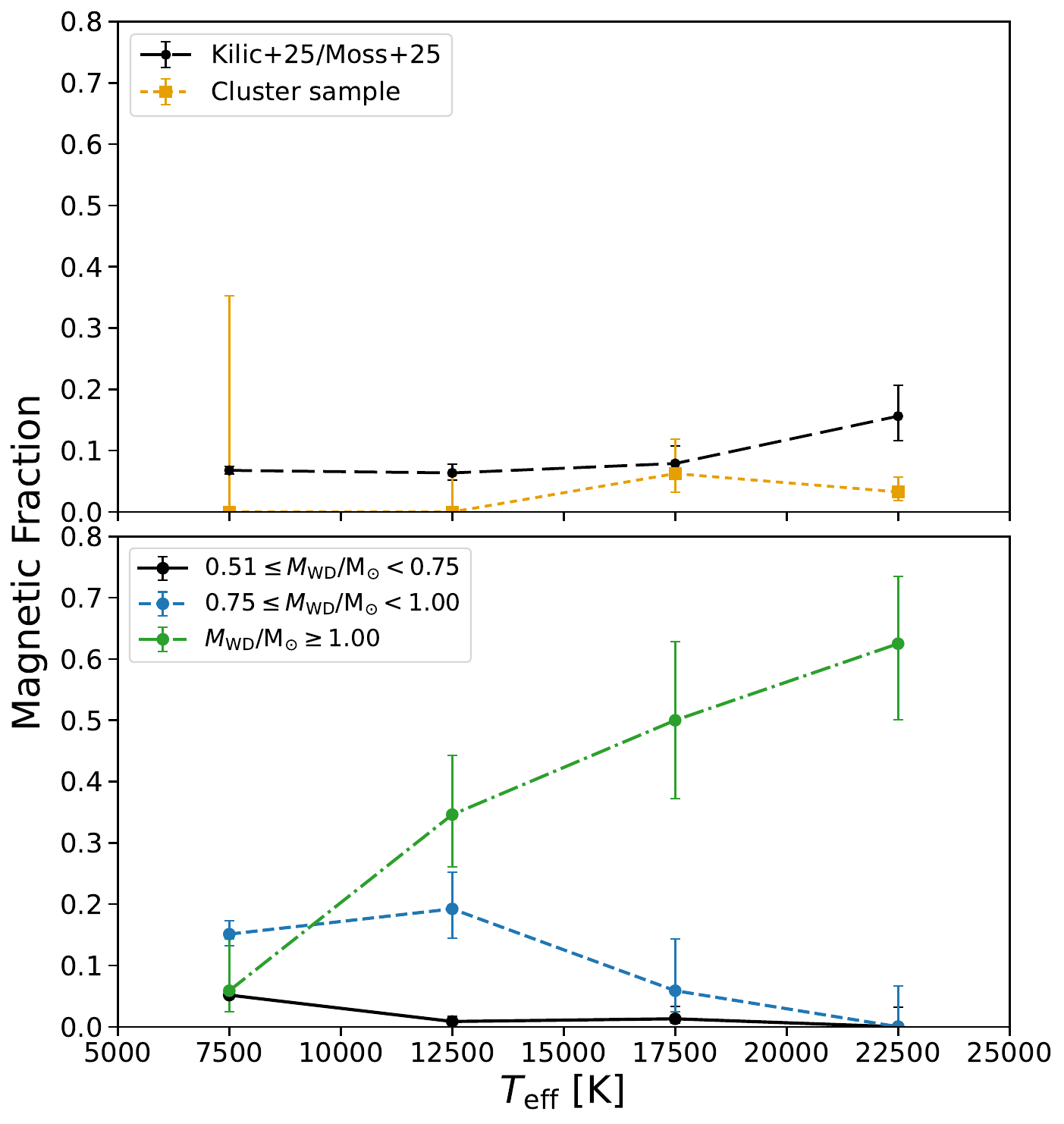}
    \caption{Magnetic fractions across temperature bins. Top: fractions for the 100~pc SDSS field sample of \citet{2025ApJ...979..157K} and \citet{2025ApJ...990...25M} and for the cluster sample. Bottom: fractions for the 100~pc SDSS field sample, separated into three WD mass ranges. Samples are restricted to DA and DB subtypes and their magnetic counterparts. Error bars show 68\% Jeffreys credible intervals.}
    \label{fig:magnetic_frac_combined}
\end{figure}

These overall fractions are not directly comparable because the two samples have very different temperature distributions. Only 64 of the 2,146 field WDs (3\%) have $T_{\rm eff}\geq20{,}000$~K, compared with 92 of 142 cluster WDs (65\%). Although magnetic WD samples extend to greater distances, we did not identify one with a similarly consistent set of WD parameters and the same focus on identifying magnetic WDs as the chosen 100~pc SDSS footprint sample. As such, we retain this sample despite the different temperature distributions. For comparison, we divide both samples into $T_{\rm eff}<10{,}000$, $10{,}000$--$15{,}000$, $15{,}000$--$20{,}000$, and $\geq20{,}000$~K bins. We group all hotter WDs into the final bin because finer subdivision leaves too few objects, while still providing comparable field and cluster sample sizes in this temperature range. The top panel of Fig.~\ref{fig:magnetic_frac_combined} shows the resulting magnetic fractions, with uncertainties given by 68\% Jeffreys credible intervals. The cluster magnetic fraction is lower in all four bins; the cooler bins contain few cluster WDs and consequently have large uncertainties, while above $20{,}000$~K the magnetic fraction is just 3.3\% in clusters compared with 15.6\% in the field.

The field magnetic fraction also shows a strong dependence on WD mass, as shown in the bottom panel of Fig.~\ref{fig:magnetic_frac_combined}. For $0.51\leq M_{\rm WD}/\msun<0.75$, magnetism is uncommon, and the fraction remains very low at all temperatures. For $0.75\leq M_{\rm WD}/\msun<1.0$, the magnetic fraction is higher at lower temperatures and declines in the hotter bins. In contrast, the magnetic fraction among WDs with $M_{\rm WD}\geq1.0\msun$ increases strongly with temperature, reaching 62.5\% above $20{,}000$~K. All 10 magnetic WDs in the field sample above $20{,}000$~K have $M_{\rm WD}\geq1.0\msun$.

\begin{figure*}[ht]
    \centering
    \includegraphics[width=\textwidth]{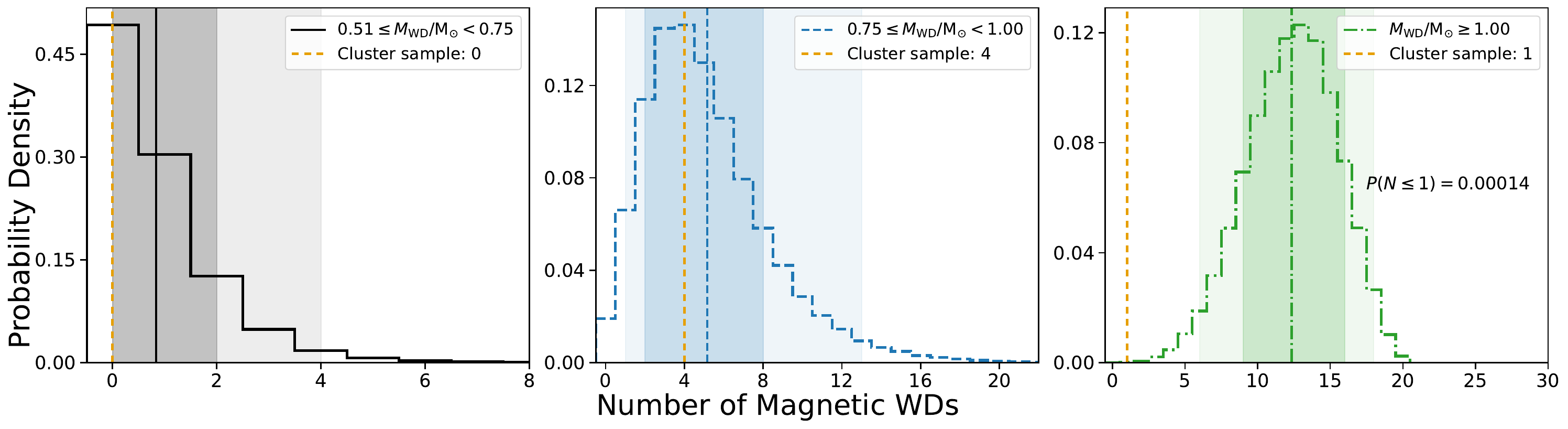}
    \caption{Expected magnetic cluster WD counts in three WD mass ranges, based on the temperature-dependent magnetic fractions of the 100~pc SDSS field sample. Shaded regions show the 68\% and 95\% posterior-predictive intervals, while vertical lines mark the predicted means and orange dashed lines the observed counts. The high-mass panel also gives $P(N\leq N_{\rm obs})$.}
    \label{fig:magnetic_expected_bymass}
\end{figure*}

We use these mass- and temperature-dependent field fractions to predict the number of magnetic WDs expected in each of the three cluster mass ranges, with the resulting posterior-predictive distributions shown in Fig.~\ref{fig:magnetic_expected_bymass}. At $0.51\leq M_{\rm WD}/\msun<0.75$, none of the 47 cluster WDs are magnetic. This is consistent with the field population, where 60 of the 63 magnetic WDs in this mass range have $T_{\rm eff}<10{,}000$~K, while only two of the 47 cluster WDs are this cool. The cluster sample contains very few WDs in the temperature range where nearly all low-mass magnetic WDs are found, so it is unsurprising that no low-mass magnetic cluster WDs are in the sample. At $0.75\leq M_{\rm WD}/\msun<1.0$, the overall magnetic incidence is consistent with the field prediction. The magnetic cluster WDs are found at somewhat higher temperatures than those in the field sample, but the individual high-temperature field bins contain few objects, and the differences are not significant. The high-mass population is markedly different. All 20 cluster WDs with $M_{\rm WD}\geq1.0\msun$ have $T_{\rm eff}\geq20{,}000$~K, yet only one is magnetic, compared with 10 of 16 field WDs in the same mass and temperature range. The field-based prediction gives a mean of 12 magnetic cluster WDs, yet only one is observed, with a posterior-predictive probability of $P(N\leq1)=1\times10^{-4}$. 

The lower magnetic incidence in clusters is therefore driven almost entirely by a strong deficit of magnetic WDs in the high-mass population. The significance of this comparison is limited by the small number of hot, high-mass field WDs, the broad $T_{\rm eff}\geq20{,}000$~K bin, and remaining differences between the field and cluster samples. Even with these limitations, the contrast is suggestive and further motivates examining the formation channels of magnetic WDs in clusters and the field. 


\subsection{Formation Channels}
\label{subsec:magnetic_formation_channels}

\begin{figure}[ht]
    \centering
    \includegraphics[width=\columnwidth]{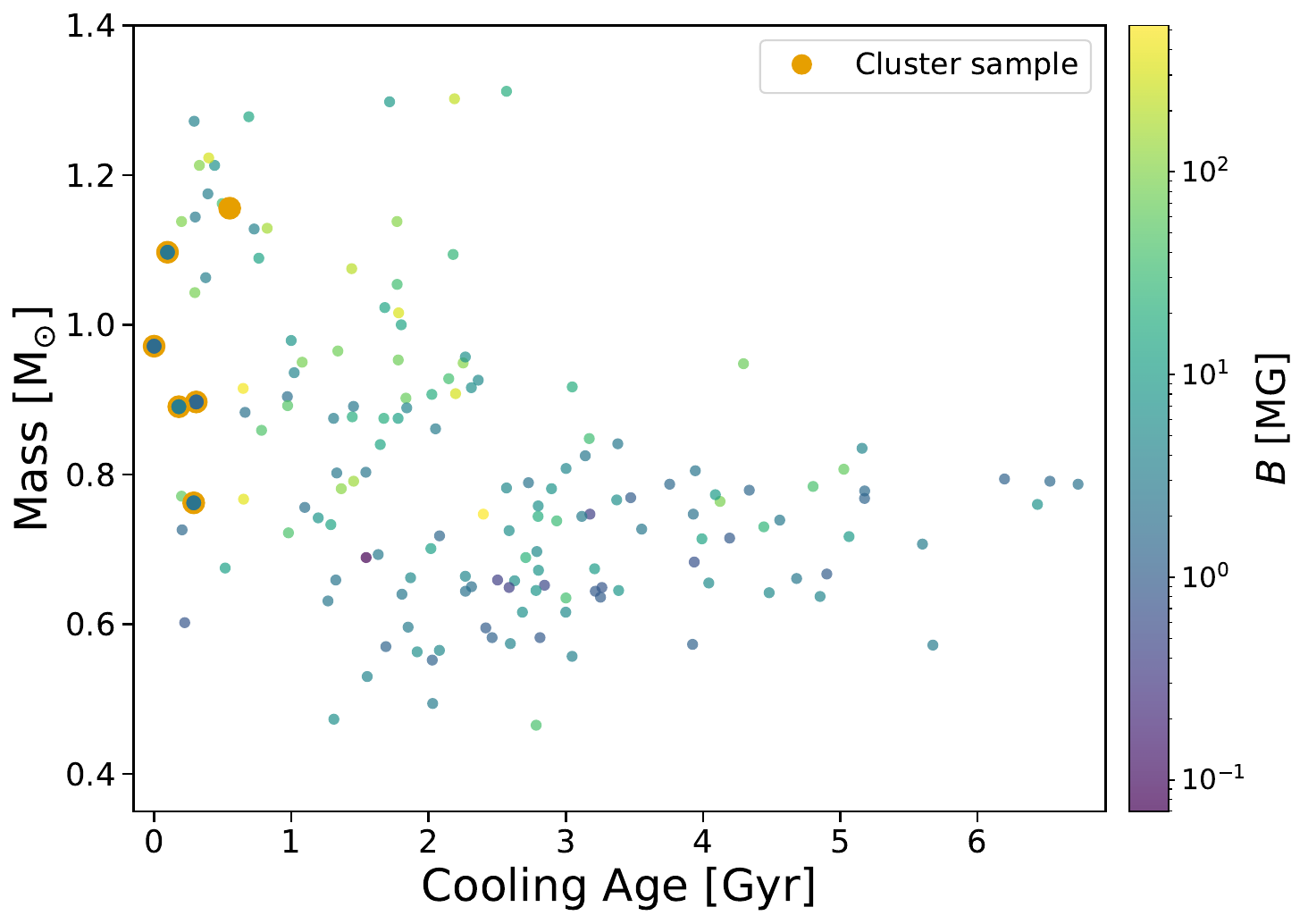}
    \caption{Magnetic cluster WDs (orange) compared with the 100~pc SDSS field population. Marker colors indicate magnetic field strength. Praesepe-WDJ082547+174818 is expected to be highly magnetic from its featureless DC spectrum, but its field strength is not well constrained.}
    \label{fig:magnetic_mass_cooling_age}
\end{figure}

Magnetic fields in WDs likely originate through multiple formation channels. Nearby field WD samples show a broad division between young, massive WDs with generally stronger fields and older, lower-mass WDs with weaker magnetism \citep{2021MNRAS.507.5902B,2022ApJ...935L..12B}. \citet{2022ApJ...935L..12B} argued that these populations likely have different dominant origins, with binary interactions contributing substantially to the young, massive population and single-star evolution dominating at lower masses. \citet{2025ApJ...990...25M} found further evidence for this division in the 100~pc SDSS sample used for our magnetic incidence comparison. Fig.~\ref{fig:magnetic_mass_cooling_age} shows our cluster magnetic WD sample alongside this field population. All six cluster WDs lie at young cooling ages, and most are relatively massive. Praesepe-WDJ082547+174818 has no magnetic field estimate because of its featureless DC spectrum, while the remaining cluster WDs have relatively weak fields. The division is not clean in field strength in the 100~pc sample, however, as lower-field WDs are also present among the young, massive field population and overlap with the cluster objects. These young cluster WDs are therefore particularly useful for probing channels that can reveal magnetic fields early in WD evolution.

One possibility is that the field originates in the WD progenitor star. Asteroseismology has revealed strong internal magnetic fields in red giants, and their dependence on stellar mass suggests an origin in dynamos in main-sequence convective cores \citep{2015Sci...350..423F,2016Natur.529..364S,2016ApJ...824...14C,2022Natur.610...43L}. Investigating whether such fields could remain buried and later reach the WD surface, \citet{2024A&A...691L..21C}
found that emergence should occur during the WD phase for masses above about $0.65\msun$, with breakout occurring earlier at higher WD mass. Extending this work, \citet{2026ApJ...998...28C} find that diffusion of surviving progenitor fields can explain many observed WD field strengths and their dependence on cooling age. Assuming equipartition in the progenitor dynamo and magnetic flux conservation, their predicted strengths generally provide an approximate upper limit to the observations. Using constraints from red giant asteroseismology, \citet{2026A&A...708L..14E} found that inherited fields can reproduce observed strengths and emergence times for magnetic WDs, provided the red giant's radiative interior is broadly magnetized. The predicted breakout times remain uncertain, particularly because of the treatment of convective boundaries and mass loss.

Crystallization offers another potential single-star formation channel. Chemical separation during crystallization can drive convection that, together with sufficiently rapid rotation, may sustain a magnetic dynamo \citep{2017ApJ...836L..28I,2022MNRAS.514.4111G}. The increased incidence of magnetism among crystallizing WDs in nearby samples provides observational support for this mechanism \citep{2021MNRAS.507.5902B,2022ApJ...935L..12B}. Recent calculations find that convection is generally too weak through most of crystallization to generate observable fields \citep{2024ApJ...961..197M}, although it has been suggested that a brief period of stronger convection at its onset may produce MG fields \citep{2024ApJ...964L..15F,2024ApJ...969...10C}. Even if a field is generated at crystallization, it may remain trapped in the convective zone and take a few Gyr to reach the WD surface \citep{2024MNRAS.528.3153B}.

The cluster sample helps constrain these two single-star channels. All five magnetic WDs with viable single-star progenitor solutions lie at cooling ages younger than even the lower uncertainty bound on the predicted main-sequence dynamo breakout times of \citet{2024A&A...691L..21C}. The hottest object has $T_{\rm eff}>100{,}000$~K and a very short WD cooling age, showing that at least some magnetic fields can be visible early in WD cooling. Crystallization occurs far too late to explain these fields, leaving an inherited or fossil field as the leading single-star origin. If their fields arise through this channel, the cluster sample suggests shorter breakout times than predicted by \citet{2024A&A...691L..21C} and may help constrain the uncertain treatments of convective boundaries and mass loss. 

M39-WDJ212659+484455, Stock2-WDJ020725+592444, and Praesepe-WDJ083945+200015 have weaker fields than predicted by \citet{2026ApJ...998...28C} at comparable masses and cooling ages, as do many field WDs in their comparison. ASCC47-WDJ083224-393543 already has an MG-strength field at a younger cooling age than their models of comparable mass predict for such strengths, while M47-WDJ073606-145131 lies above the model mass range. \citet{2026A&A...708L..14E} find that inherited fields can explain their WD observations if the field extends beyond the former main-sequence convective core. This suggests that explaining the early magnetism in our sample may require more extended progenitor fields, although their models would need to be extended to higher WD masses to test this possibility.

\begin{figure}[ht]
    \centering
    \includegraphics[width=\columnwidth]{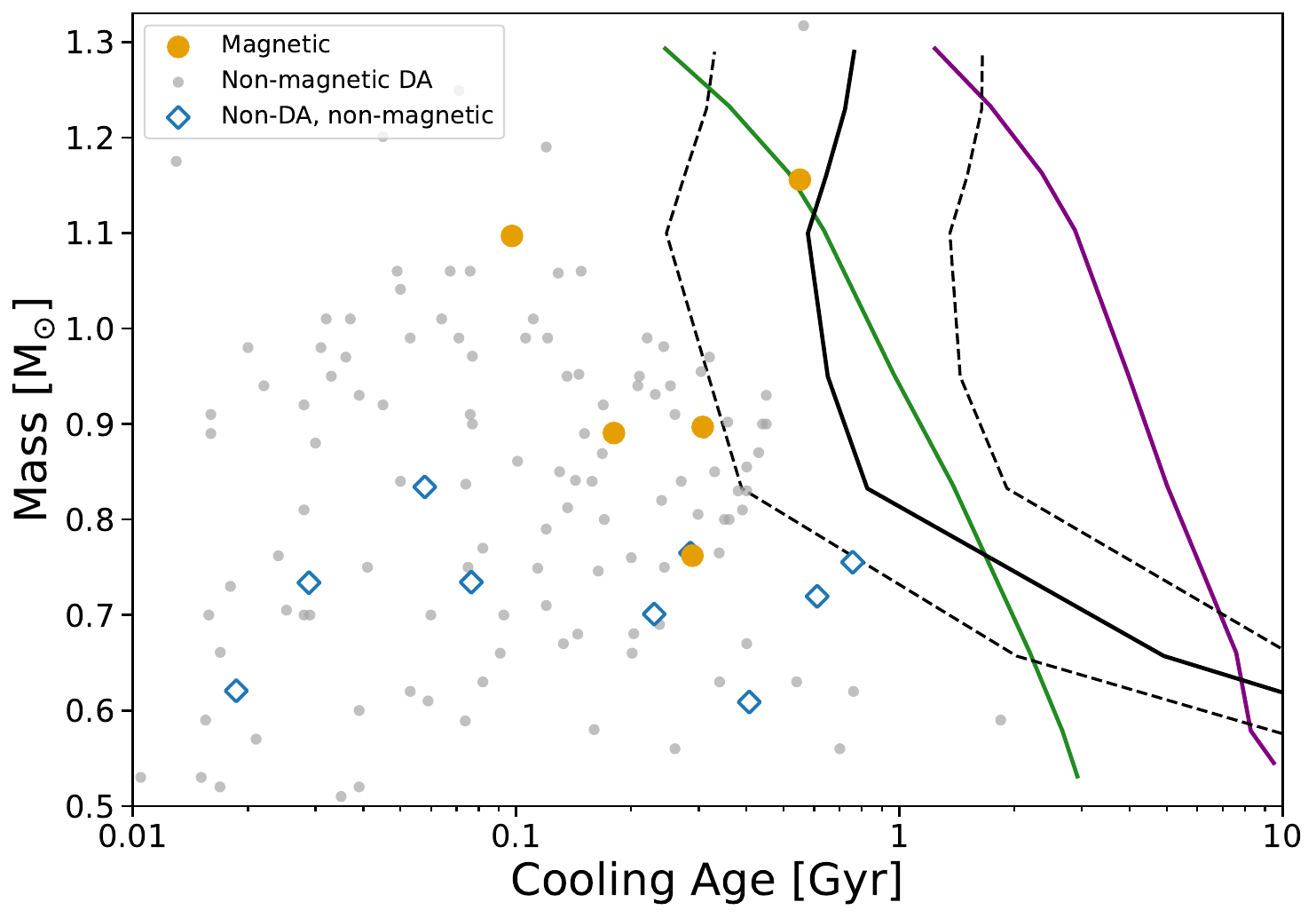}
    \caption{Mass and cooling age distribution of the cluster WD sample. The green and purple curves show the onset of crystallization and crystallization-driven field emergence, respectively, while the solid and dashed black curves show estimated main-sequence dynamo field emergence times and its uncertainty bounds \citep{2024A&A...691L..21C}. One magnetic WD with $t_{\rm cool}<0.01$~Gyr lies outside the plotted range.}
    \label{fig:mass_cooling_age}
\end{figure}

The broader cluster population provides a complementary constraint. As shown in Fig.~\ref{fig:mass_cooling_age}, the sample contains many lower-mass WDs, yet only one magnetic WD below $0.88\msun$. At these masses, the magnetic incidence in the field is higher than in the cluster sample. All of the lower-mass cluster WDs lie before the predicted onset of crystallization, and most are also younger than the nominal emergence times for fields generated by main-sequence dynamos \citep{2024A&A...691L..21C}. Together, these results suggest that field emergence can occur rapidly in some higher-mass WDs while remaining delayed at lower masses, where both crystallization and later emergence of progenitor fields may contribute to the magnetic population.

Binary interactions also contribute substantially to the magnetic population, with strong fields potentially generated during double WD mergers or common envelope evolution \citep{2008MNRAS.387..897T,2012ApJ...749...25G}. Observational evidence for a merger contribution is strongest among massive WDs. Young, massive WDs in the 40~pc sample are frequently magnetic and tend to have stronger fields \citep{2022ApJ...935L..12B}. A substantial merger contribution to the massive WD population is also inferred independently from kinematic and cooling age distributions \citep{2020ApJ...891..160C,2022MNRAS.511.5984F}. Similar, \citet{2024ApJ...974...12J} find high merger fractions among massive WDs, many of which show rapid rotation or unusual kinematics commonly associated with merger remnants. These results are consistent with the young, massive magnetic population seen in the 100~pc sample of \citet{2025ApJ...990...25M}.

Cluster membership provides an important constraint on binary formation channels. As shown in Fig.~\ref{fig:magnetic_ifmr}, the five magnetic WDs in this comparison all lie within $1\sigma$ of the best-fit IFMR from \miller. Their viable progenitor masses, agreement with the IFMR, and relatively weak magnetic fields together support single-star origins. A double WD merger scenario is also strongly limited by the cluster ages and observed WD masses. Following the argument of \citet{2020ApJ...901L..14C}, the relatively massive progenitors required to form two WDs within these young clusters make it difficult to produce merger remnants at the observed WD masses. Uncertainties remain in cluster ages, WD cooling models, and the form of the IFMR, and a merger could in principle reset the cooling clock and still leave a remnant near the observed relation \citep{2020A&A...636A..31T}. Even so, these constraints strongly favor single-star origins for these five magnetic WDs.

The high-mass magnetic cluster population provides an additional constraint on merger channels. All 20 cluster WDs with $M_{\rm WD}\geq1.0\msun$ also have $T_{\rm eff}\geq20{,}000$~K, yet only one is magnetic, compared with 10 of 16 field WDs in the same mass and temperature range. The cluster sample therefore shows that young, massive WDs can form through apparently normal single-star evolution without developing magnetism at the high rate seen in the field. The difference may reflect the time available for merger channels. Although some merger pathways occur on relatively short timescales, many produce remnants only after substantial delays, and the cluster sample is heavily weighted to younger systems, with just eight of the 41 clusters represented in the combined sample older than 1~Gyr. Older clusters provide more time for such channels, but also progressively lose members and may dissolve through dynamical evolution \citep{2003MNRAS.338..717B}. The field can therefore accumulate delayed merger remnants over much longer evolutionary histories, which may help explain the much higher observed magnetic incidence at high WD mass.

Overall, the cluster sample points to the strong role evolutionary time plays in the appearance of WD magnetism. At lower masses, the lack of magnetism among young cluster WDs is consistent with fields becoming visible only later through crystallization or delayed emergence from the progenitor. In contrast, the five magnetic WDs with viable single-star progenitor solutions show that some fields can be visible much earlier than the breakout times predicted by \citet{2024A&A...691L..21C}, including essentially from the start of WD cooling in the hottest object. At high mass, the cluster sample shows that massive WDs can form through normal single-star evolution without developing magnetism at the high rate seen in the field. This difference may reflect the field accumulating magnetic merger remnants over longer evolutionary timescales, while most clusters in our sample are too young to host comparable populations. 


\subsection{Praesepe-WDJ082547+174818}
\label{subsec:praesepe_dc_wd}

\begin{figure}[ht]
    \centering
    \includegraphics[width=\columnwidth]{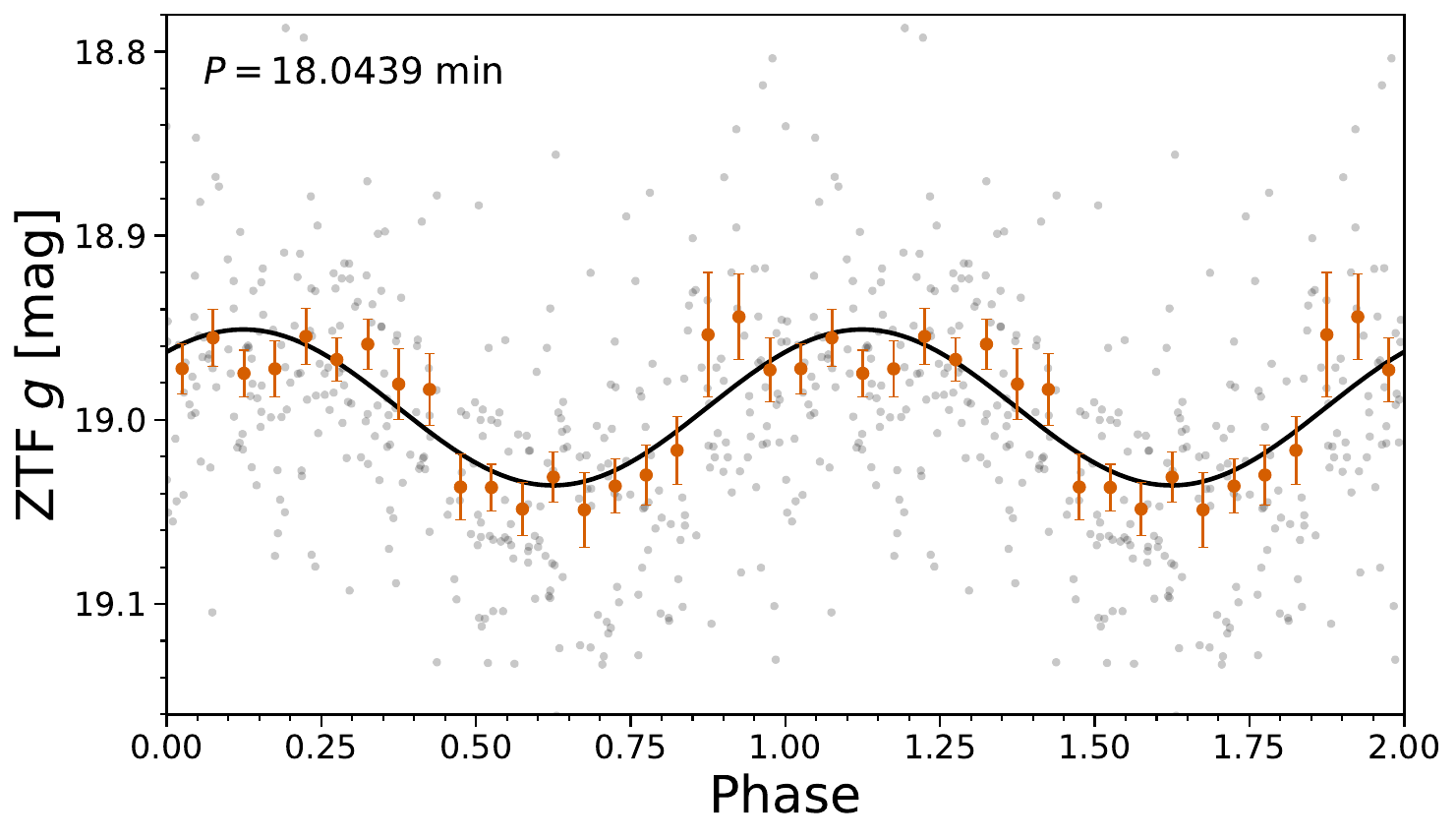}
    \caption{Phase-folded ZTF $g$-band photometry of Praesepe-WDJ082547+174818 at $P=18.0439$~min. Orange points show the phase-binned mean photometry, and the black curve shows the best-fitting sinusoid.}
    \label{fig:praesepe_dc_ztf}
\end{figure}

Praesepe-WDJ082547+174818 is particularly noteworthy because its optical light curve in the Zwicky Transient Facility archive \citep[ZTF,][]{2019PASP..131a8003M,2019PASP..131a8002B,2019PASP..131g8001G,2020PASP..132c8001D} shows an 18-minute periodic signal, as shown in Fig.~\ref{fig:praesepe_dc_ztf}. As discussed in Section~\ref{subsubsec:photometric_fitting}, we interpret its featureless spectrum as evidence for a strong magnetic field and pragmatically adopt the DA solution from the SED fit. The source lies within the footprints of 2MASS and ALLWISE but has no infrared counterpart, providing no evidence for an IR excess or luminous cool companion. 

The combination of strong magnetism, an 18-minute periodic signal, and the absence of evidence for a luminous cool companion suggests that this source may be a rapidly rotating magnetic WD, with the variability produced by rotational modulation \citep[e.g.,][]{2004MNRAS.348L..33B,2019MNRAS.489.3648K}. The infrared non-detections make variability associated with a luminous companion unlikely, although they do not rule out a faint brown dwarf companion, and the photometry alone does not establish rotation. Rapid rotation is commonly associated with magnetic merger remnants \citep[e.g.,][]{2021Natur.595...39C,2022AJ....164..131W,2026A&A...706A.188C,2026A&A...712A.230D}, making a past merger a plausible origin for this object. Dedicated time-resolved photometry would be needed to confirm the signal independently and establish whether it traces the WD's rotation.

The adopted solution yields a WD mass of $1.156\pm0.007\msun$ and a cooling age of $550^{+50}_{-40}$~Myr. Using the \miller best-fit IFMR, a single star producing a WD of this mass would have an initial mass of approximately $6.83\msun$, corresponding to a PARSEC progenitor lifetime of approximately $55$~Myr at the adopted Praesepe metallicity. Together with the cluster age, this yields an estimated evolutionary delay of $136^{+82}_{-58}$~Myr relative to single-star formation. The quoted uncertainty combines the cluster age, cooling age, and WD mass uncertainties assuming independence and a fixed IFMR; it excludes IFMR scatter and systematic uncertainties in the adopted non-magnetic models. 


\section{Summary and Conclusions}
\label{sec:conclusions}

Using an extensive literature census supplemented by candidate atypical WDs identified in the archival spectroscopic census of \citet{Miller2026Surveys}, we examine the implications of the known atypical WD population in open clusters. We identify 19 candidate atypical cluster WDs, one of which is subsequently ruled out as a cluster member, leaving a final sample of 18: eight non-magnetic DB/DBA WDs, five DC WDs, and five magnetic DA/DB WDs. For these objects, we derive consistent cluster and WD parameters, including progenitor lifetimes and masses where a single-star interpretation is favored. By combining this sample with the larger cluster DA population, we use the additional constraints from cluster membership to compare atypical WDs in clusters and the field, examine their evolutionary origins, and assess how the cluster environment influences the observed populations.

For the nine non-DAs with viable single-star progenitor solutions, we find little evidence that they differ from the DA population. The mean offset from the DA IFMR is $\Delta M_f=-0.035\pm0.037\msun$, consistent with zero. We likewise find no robust spatial or kinematic differences between the cluster DA and non-DA populations. Both are more centrally concentrated than their host-cluster members, though this may partly reflect spectroscopic follow-up selection bias.

The most striking result, however, is that two independent properties of the high-mass WD population point to the same difference between clusters and the field. First, clusters lack the massive non-DA population seen in the field, particularly DQs and other peculiar WDs. The strongest discrepancy occurs at $0.8\leq M_{\rm WD}/\msun<1.0$, where the field population predicts $6.0^{+0.8}_{-0.7}$ non-DAs but none are observed, with a posterior-predictive probability of $P(N=0)=3.7\times10^{-3}$. Second, massive magnetic WDs are similarly rare: only one of 20 cluster WDs with $M_{\rm WD}\geq1.0\msun$ and $T_{\rm eff}\geq20{,}000$~K is magnetic, compared with 10 of 16 field WDs, with $P(N\leq1)=1\times10^{-4}$. At lower masses, the cluster magnetic incidence is broadly consistent with the field once we account for different temperature distributions. Together, these results suggest that open clusters are deficient in the massive WD populations associated with merger formation, while the incidence of DBs is similar to the field.

The simplest explanation for the apparent lack of merger remnants is the different evolutionary timescales sampled by clusters and the field. Most clusters contributing WDs to the present sample are young, leaving relatively little time for delayed merger remnants to accumulate. The Galactic field, in contrast, accumulates merger remnants over much longer timescales. Open clusters old enough to produce substantial delayed merger populations are also increasingly likely to lose members or dissolve, further suppressing these objects in present-day cluster samples. Open clusters may therefore show the high-mass WD population before delayed mergers contribute substantially, whereas the field contains the cumulative products of both single-star and binary evolution.

Five magnetic cluster WDs strongly favor single-star origins. Their young cooling ages suggest that fossil fields may emerge earlier than the breakout times estimated by \citet{2024A&A...691L..21C}, helping constrain mass loss and the treatment of convective boundaries. Crystallization occurs far too late to explain these objects, while the lack of magnetism among young lower-mass cluster WDs remains consistent with later emergence through crystallization or delayed breakout. Praesepe-WDJ082547+174818 is also a likely highly magnetic WD whose 18.0439-minute ZTF period may indicate rapid rotation following a past merger, although time-resolved observations are needed to confirm this interpretation.

Overall, the expanded sample of atypical cluster WDs reveals significant differences from the field in both the high-mass non-DA and magnetic populations. While the current sample remains limited by small numbers and uneven coverage in mass and temperature, larger spectroscopic samples will strengthen these comparisons and, combined with the age and progenitor constraints provided by cluster membership, enable increasingly stringent tests of WD evolution.


\begin{acknowledgments}

This work was supported in part by the Natural Sciences and Engineering Research Council of Canada Discovery Grants DG-RGPIN-2022-03051 and DG-RGPIN-2023-04486. This research received funding from the European Research Council under the European Union's Horizon 2020 research and innovation program number 101002408 (MOS100PC).

This work has made use of data from the European Space Agency (ESA) mission {\it Gaia} (\url{https://www.cosmos.esa.int/gaia)}, processed by the {\it Gaia} Data Processing and Analysis Consortium (DPAC, \url{https://www.cosmos.esa.int/web/gaia/dpac/consortium}). Funding for the DPAC has been provided by national institutions, in particular the institutions participating in the {\it Gaia} Multilateral Agreement. 

The Pan-STARRS1 Surveys (PS1) and the PS1 public science archive have been made possible through contributions by the Institute for Astronomy, the University of Hawaii, the Pan-STARRS Project Office, the Max-Planck Society and its participating institutes, the Max Planck Institute for Astronomy, Heidelberg and the Max Planck Institute for Extraterrestrial Physics, Garching, The Johns Hopkins University, Durham University, the University of Edinburgh, the Queen's University Belfast, the Harvard-Smithsonian Center for Astrophysics, the Las Cumbres Observatory Global Telescope Network Incorporated, the National Central University of Taiwan, the Space Telescope Science Institute, the National Aeronautics and Space Administration under Grant No. NNX08AR22G issued through the Planetary Science Division of the NASA Science Mission Directorate, the National Science Foundation Grant No. AST-1238877, the University of Maryland, Eotvos Lorand University (ELTE), the Los Alamos National Laboratory, and the Gordon and Betty Moore Foundation. 

This work includes data collected by the Galex mission which is publicly available from the Mikulski Archive for Space Telescopes (MAST, \url{https://archive.stsci.edu/}). 

Based on data products from observations made with ESO Telescopes at the La Silla Paranal Observatory under programme ID 177.D-3023, as part of the VST Photometric H$\alpha$ Survey of the Southern Galactic Plane and Bulge (VPHAS+, \url{www.vphas.eu}). 

Funding for the Sloan Digital Sky Survey V has been provided by the Alfred P. Sloan Foundation, the Heising-Simons Foundation, the National Science Foundation, and the Participating Institutions. SDSS acknowledges support and resources from the Center for High-Performance Computing at the University of Utah. SDSS telescopes are located at Apache Point Observatory, funded by the Astrophysical Research Consortium and operated by New Mexico State University, and at Las Campanas Observatory, operated by the Carnegie Institution for Science. The SDSS web site is \url{www.sdss.org}. SDSS is managed by the Astrophysical Research Consortium for the Participating Institutions of the SDSS Collaboration, including the Carnegie Institution for Science, Chilean National Time Allocation Committee (CNTAC) ratified researchers, Caltech, the Gotham Participation Group, Harvard University, Heidelberg University, The Flatiron Institute, The Johns Hopkins University, L'Ecole polytechnique f\'{e}d\'{e}rale de Lausanne (EPFL), Leibniz-Institut f\"{u}r Astrophysik Potsdam (AIP), Max-Planck-Institut f\"{u}r Astronomie (MPIA Heidelberg), Max-Planck-Institut f\"{u}r Extraterrestrische Physik (MPE), Nanjing University, National Astronomical Observatories of China (NAOC), New Mexico State University, The Ohio State University, Pennsylvania State University, Smithsonian Astrophysical Observatory, Space Telescope Science Institute (STScI), the Stellar Astrophysics Participation Group, Universidad Nacional Aut\'{o}noma de M\'{e}xico, University of Arizona, University of Colorado Boulder, University of Illinois at Urbana-Champaign, University of Toronto, University of Utah, University of Virginia, Yale University, and Yunnan University. 

The national facility capability for SkyMapper has been funded through ARC LIEF grant LE130100104 from the Australian Research Council, awarded to the University of Sydney, the Australian National University, Swinburne University of Technology, the University of Queensland, the University of Western Australia, the University of Melbourne, Curtin University of Technology, Monash University and the Australian Astronomical Observatory. SkyMapper is owned and operated by The Australian National University's Research School of Astronomy and Astrophysics. The survey data were processed and provided by the SkyMapper Team at ANU. The SkyMapper node of the All-Sky Virtual Observatory (ASVO) is hosted at the National Computational Infrastructure (NCI). Development and support of the SkyMapper node of the ASVO has been funded in part by Astronomy Australia Limited (AAL) and the Australian Government through the Commonwealth's Education Investment Fund (EIF) and National Collaborative Research Infrastructure Strategy (NCRIS), particularly the National eResearch Collaboration Tools and Resources (NeCTAR) and the Australian National Data Service Projects (ANDS). 

Based on observations obtained with the Samuel Oschin Telescope 48-inch and the 60-inch Telescope at the Palomar Observatory as part of the Zwicky Transient Facility project. ZTF is supported by the National Science Foundation under Grants No. AST-1440341 and AST-2034437 and a collaboration including current partners Caltech, IPAC, the Oskar Klein Center at Stockholm University, the University of Maryland, University of California, Berkeley, the University of Wisconsin at Milwaukee, University of Warwick, Ruhr University, Cornell University, Northwestern University and Drexel University. Operations are conducted by COO, IPAC, and UW. 

This paper makes use of data obtained as part of the IGAPS merger of the IPHAS and UVEX surveys (\url{www.igapsimages.org}) carried out at the Isaac Newton Telescope (INT).
\end{acknowledgments}

\software{Astropy \citep{2013A&A...558A..33A,2018AJ....156..123A,2022ApJ...935..167A},
pyphot \citep{fouesneau2025pyphot}}

\facilities{Gaia (DR3), GALEX, PS1, Sloan, ZTF}

\section*{Data Availability}
Data for the 123 previously published DAs in the comparison sample are available at \url{https://wdifmr.org}. The nine addtional DAs analyzed in \citet{Miller2026Surveys} will be added upon acceptance of that manuscript.


\bibliographystyle{aasjournalv7.1}
\bibliography{main}



\end{document}